\def\ba {\begin{array}}
\def\ea {\end{array}}
\def\bea {\begin{eqnarray}}
\def\eea {\end{eqnarray}}
\def\e{{\rm e}}
\def\bl#1{{\overline {l_{#1}}}}
\def\b{{\frak b}}
\def\bb{\overline{{\frak b}}}
\def\c{{\frak c}}
\def\bc{\overline{{\frak c}}}
\def\t{\tau}
\def\bt{\overline{\tau}}
\def\B{{\frak B}}
\def\bB{\overline{{\frak B}}}
\def\C{{\frak C}}
\def\bC{\overline{{\frak C}}}
\def\T{T}
\def\bT{\overline{T}}
\def\squareforqed{\hbox{\rlap{$\sqcap$}$\sqcup$}}
\def\qed{\ifmmode\squareforqed\else{\unskip\nobreak\hfil
\penalty50\hskip1em\null\nobreak\hfil\squareforqed
\parfillskip=0pt\finalhyphendemerits=0\endgraf}\fi}

\def\wcirc{\squareforqed}
\def\Im{{\rm Im}}

\def\L#1{${\cal L}_#1$}
\def\sing{\equiv_s}
\def\<{\langle}
\def\>{\rangle}
\documentstyle[12pt,amsmath,amsfonts,amssymb,graphicx]{article}

\begin{document}
\title{The Hubbard chain at finite temperatures: ab initio calculations
of Tomonaga-Luttinger liquid properties}
\author{G.~J\"uttner\thanks{e-mail: gj@thp.uni-koeln.de},        
        A. Kl\"umper\thanks{e-mail: kluemper@thp.uni-koeln.de},
        J. Suzuki\thanks{e-mail: suz@hep1.c.u-tokyo.ac.jp,
        Permanent address: Institute of Physics, 
                           University of Tokyo at Komaba}\\
        \parbox{0.9\textwidth}{
        {\em
        \begin{center}
        Universit\"at zu K\"oln  \\                
        Institut f\"ur Theoretische Physik \\     
        Z\"ulpicher Str. 77  \\                   
        D-50937, Germany
        \end{center}
        }}  
       }    

\date{October 1997}

\maketitle
\begin{abstract}
We present a novel treatment of finite temperature properties
of the one-dimensional Hubbard model. Our approach is based on a 
Trotter-Suzuki mapping utilizing Shastry's classical model and
a subsequent investigation of the quantum transfer matrix. 
We derive non-linear integral equations for three auxiliary functions which
have a clear physical interpretation of elementary excitations of spin type
and charge excitations in lower and upper Hubbard bands. This
allows for a transparent analytical study of certain limiting cases
as well as for precise numerical investigations. We present data
for the specific heat, magnetic and charge susceptibilities for
various particle densities and coupling strengths $U$. The structure
exposed by these curves is discussed in terms of the elementary 
charge and spin excitations. Special emphasis is placed on the study  
of the low-temperature behavior within our ab initio approach confirming 
the scaling predictions by Tomonaga-Luttinger liquid theory. 
In addition we make 
contact with the ``dressed energy'' formalism established for the 
analysis of ground state properties.
\end{abstract}

\clearpage
\section{Introduction}

The Hubbard model represents the most fundamental model for
highly correlated electron systems. It has therefore attracted
much attention since its formulation.
Triggered by the discovery of high $T_c$ superconductivity,
correlation effects of the model have been of strong recent interest.

For the problem in one spatial dimension an exact solution is 
available via the Bethe ansatz \cite{LiebWu68}. 
It clarified the Mott insulator nature of the itinerant electron
system at half-filling in 1D.
The extensive list of investigations of zero temperature properties 
ranges from studies of the elementary excitations \cite{Ov70,Coll74,Woy82,
KlSchZ90},
over magnetic properties \cite{Shi72b} to studies of correlation
functions in the strong coupling limit \cite{OgaShi90}. Recently, the large 
group of symmetries of the Hubbard model was analyzed in \cite{EssKorSch91,
UgKor94,Mur97,Yang89}.
It was, however, almost 20 years after the discovery of the exact solution
that various properties like asymptotics of correlations functions
at $T=0$ have been investigated by using results
of conformal field theory.
The Tomonaga-Luttinger liquid properties of the one-dimensional
model have been shown not only at a
qualitative but also at satisfactory quantitative level
\cite{FraKor90,FraKor91,Woyn89,Schu91}.

In this report, we address the problem of the Hubbard model
at finite temperatures. 
In fact, soon after the seminal solution \cite{LiebWu68},
a thermodynamic formulation  
was set up through the string hypothesis approach \cite{Tak72}.
It results into infinitely many ($\infty \times \infty$) coupled
nonlinear integral equations for infinitely many unknown
functions.
Obviously, quantitative studies of such equations need
much effort and were performed only relatively recently (also
about 20 years after the thermodynamical formulation!)
\cite{KawUsuOki89,UsuKawOki90}.
Still,  the explicit calculation allowed for
only 2 (!)   bound charge rapidities and  15-30 spin rapidities, while
the  original equations  contain $\infty  \times  \infty$  rapidities.
%

Here we attack the problem via a completely different approach
developed recently which avoids the computational complications and
also renders correlations lengths at finite temperatures accessible.
We make use of a general equivalence theorem  between 
$d$-dimensional quantum systems at  finite
temperatures and $d+1$-dimensional classical systems \cite{Suz85}, 
thus we employ a convenient mapping  to a two-dimensio\-nal  classical
model.
The evaluation of the free energy then reduces to 
finding the largest eigenvalue of the so-called 
quantum transfer matrix (QTM) 
\cite{Suz85,SuIn87,Koma87,Suz90,Tsune91,Tak91,Klu92,Klu93,DesVeg92,SNOW92}.
The crucial observation is the existence of a commuting family of
QTMs labeled by one complex parameter (spectral parameter)
\cite{Klu92,Klu93,JutKlu96,JutKluSuz97,JutKluSuz97b}.
This makes the meaning of integrability manifest, and allows
for the investigation of thermodynamics through the
study of the analytical properties of suitably defined auxiliary functions
on the complex space of the spectral parameter.
One of the most practical advantages in this novel formulation is
the fact that one has to deal with only a finite number of 
auxiliary functions and nonlinear integral equations among them.
Therefore, we can expect results with higher numerical precision.
Furthermore, the involved auxiliary functions have
clear physical interpretations in terms of elementary excitations
at $T=0$.

Such a strategy has been successfully applied to several interesting 
models including the spin 1/2 XYZ model and derivates
\cite{Klu92,Klu93,SNOW92,DesVeg92}, 
the integrable $t-J$ model \cite{JutKlu96,JutKluSuz97},
the supersymmetric $U$ model \cite{JutKluSuz97b}, and (with limited success)
to the Hubbard model \cite{Bar81,Tsune91,KluBar96}.
The main restriction of the previous work on the Hubbard chain
\cite{Tsune91,KluBar96} is the limitation
to the case of half-filling. For any finite doping the resultant equations
appeared to be numerically ill-posed. Here we want to overcome the technical 
difficulties and derive a set of equations that allow 
for convenient (numerical) studies and clear analytical insight
for all particle densities. Before doing so, we want to remind of the
remarkable differences in comparison to other solvable models. 
The $R-$matrix for the classical analogue (``Shastry's model'') 
\cite{Shas88,OlWadAk87,WadOlAk87}
 does not possess the difference property of rapidities. 
Therefore, the intertwiner depends on two spectral parameters, not 
only on the difference.
Such violations are only known for Shastry's model and
the chiral Potts model. 
In view of analyticity, the Hubbard model is also quite
unique.
One can easily recognize this by comparing the Bethe
ansatz equations (BAE) at $T=0$ for the Hubbard model \cite{LiebWu68},
with, for instance, those for the integrable $t-J$ model 
\cite{Suth75,Sch87,BarBlaOg91}.
In both cases there are two kinds of BAE roots corresponding to charge
and spin degrees of freedom. However
for the integrable $t-J$ model both types of roots vary from $-\infty$ to
$\infty$, while the charge-rapidities for the Hubbard model only vary 
from $-\pi$ to $\pi$ with a corresponding periodicity.
This different character of the BAE roots
brings about branch cuts in a complex parameter plane 
as we will see below.
The roots show an exotic behavior:
they flow from one Riemann sheet to the other 
with changing temperature.

Despite these difficulties, we will show our strategy is
successfully applicable to the Hubbard model (and finally overcomes the
technical problems still left in the earlier approach \cite{KluBar96}).
With a careful choice of auxiliary functions, we can
completely encode the information about zeros in both sheets.
They are shown to have close relation to the physical
excitations of holons and spinons in the $T\rightarrow 0$ limit.
This limit will be studied in quite some detail as it allows for a
first principles derivation of Tomonaga-Luttinger liquid properties at low 
but finite temperatures.
Several quantities of physical interests are
evaluated with high numerical precisions for wide ranges of
temperatures and fillings.

This paper is organized as follows. In Section 2 we present the mapping
of the Hubbard chain at finite temperature to a two-dimensional classical
system with integrable QTM. In Section 3 the eigenvalue equations for the
QTM are derived which are cast into a difference type form in Section 4.
Section 5 is devoted to the derivation of non-linear integral equations
for Hubbard interactions $U>0$.
In Section 6 the integral equations are investigated numerically and the
results are discussed. Section 7 deals with the analytical study of various 
limiting cases, notably the low-temperature asymptotics. In Section 8 we
present our summary and outlook. The derivation of the integral form of 
the QTM eigenvalue is deferred to Appendix A.

\section{Shastry's Model as a classical analogue of the 1D Hubbard Model}

In the novel formalism, it is essential to deal with the
two dimensional classical counterpart.
Fortunately,  Shastry has already found 
two classical versions for the Hubbard model \cite{Shas86,Shas88}.
For the latter model,
a proof of the Yang-Baxter integrability 
has been given in a recent analysis \cite{ShiWad95} by the use of
the tetrahedron algebra. 
As the Yang-Baxter relation makes the 
finite temperature analysis much easier,
we adopt the latter version here.

We sketch the essential properties of the model.
The Hubbard model describes a lattice fermion system
with electron hopping term and 
on-site Coulomb repulsion with Hamiltonian
\begin{eqnarray}
{\cal H}_{{\rm Hubbard}} &=&
 \sum_{i=1}^L 
       \left\{ \sum_{\sigma=\pm}-( c^{\dagger}_{i+1,\sigma} c_{i,\sigma} +
     c^{\dagger}_{i,\sigma} c_{i+1,\sigma} ) 
      + U(n_{i,-} -{\textstyle \frac{1}{2}})
          (n_{i,+}-{\textstyle \frac{1}{2}}) \right\} \nonumber\\
 &+& {\cal H}_{{\rm external}}.
\end{eqnarray}
The external field term 
${\cal H}_{{\rm external}}=
-\sum_{i} [\mu ( n_{i,+}+n_{i,-})+H/2(n_{i,+}-n_{i,-})]$
will be omitted for the time being.
According to \cite{Shas88}, it is easier to find a classical
analogue after  performing the Jordan-Wigner transformations
for electrons in 1D.
The resultant spin Hamiltonian is 
\begin{eqnarray}
{\cal H}_L &=& \sum_{n=1}^{L} {\cal H}_{n, n+1}, \nonumber\\
{\cal H}_{n, n+1} &=& 
      ( \sigma^+_n  \sigma^-_{n+1} + \sigma^+_{n+1}  \sigma^-_{n})+
          ( \tau^+_n  \tau^-_{n+1} + \tau^+_{n+1}  \tau^-_{n}) +
          {\textstyle \frac{U}{4}}\, \sigma_n^z \tau_n^z,
\end{eqnarray}
where $L$ denotes the chain length of the system.
Note that we are now imposing periodic boundary conditions 
for the spin system
($\sigma_1 (\tau_1) = \sigma_{L+1} (\tau_{L+1})$).
This does not give the periodic boundary conditions for the
underlying electron system. 
The differences in boundary conditions, however, will not affect
thermodynamic quantities like the specific heat.

For the counterpart in two dimensions, 
we consider two double-layer square lattices, say a 
$\sigma$  and a $\tau$ lattice.
Each edge possesses a local variable $\pm$ and each vertex
satisfies the ice rule. 
The vertex weights consist of contributions from both
intra and inter lattice interactions.
The intra part is given by the product of vertex weights 
of the free-fermion six vertex model:
$\ell_{1,2} (u) = \ell^{\sigma}_{1,2}(u) \otimes  \ell^{\tau}_{1,2}(u)$
where
\begin{equation}
\ell^{\sigma}_{1,2}(u) = \frac{a(u)+b(u)}{2}+ 
\frac{a(u)-b(u)}{2} \sigma^z_1 \sigma^z_2 +
c(\sigma_1^+ \sigma_2^- + \sigma_1^- \sigma_2^+)
\end{equation}
and $a(u)=\cos(u)$, $b(u)=\sin(u)$, $c(u)=1$.
Taking account of inter-layer interactions, 
the following  local vertex weight
operator (denoted by $S$) is found \cite{Shas88} 
\begin{eqnarray*}
S_{1,2}(v,u) 
&=& \cos(u+v)\, \cosh(h(v,U)-h(u,U))\, \ell_{1,2} (v-u) \\
&+&  \cos(v-u)\, \sinh(h(v,U)-h(u,U))\, \ell_{1,2} (u+v)\, \sigma_2^z \tau_2^z
\end{eqnarray*} 
where $\sinh 2 h(u,U) := U a(u) b(u)/2 $. 
The Yang-Baxter relation for triple $S$ matrices is proved in 
\cite{ShiWad95}.
The commutativity of the row-to-row transfer matrix,
\begin{equation}
{\cal T}(u) := \prod^{\leftarrow}_i S_{i, g} (u,0)
\end{equation}
is the direct consequence. 

The $S$ matrix and ${\cal H}$ are related by
the expansion in small spectral parameters,
\begin{equation}
S_{1,2} (u,0)=S_{1,2}(0,-u) \sim P( 1+u {\cal H}_{1,2} + O(u^2)),
\end{equation}
where $P$ denotes the permutation operator, $P(x\otimes y) = y\otimes x$.
Once the Yang-Baxter relation is established, we can apply our machinery 
for thermodynamics. 
Here we summarize the necessary formulas, see \cite{JutKluSuz97} for details.
We define $R_{1,2}(u,v) =  S_{1,2}(v,u)|_{U \rightarrow -U}$,
and introduce $\widetilde{R}(u,v)$  and $\overline{R}(u,v)$
by clockwise and anticlockwise 
90$^\circ$ rotations of $R_{1,2}(u,v)$.
We further introduce an auxiliary transfer matrix  $\overline{\cal T}(u)$
made of Boltzmann weights $\overline{R}(0,-u)$.
The partition function is given by
\begin{equation}
Z =  \lim_{L\rightarrow \infty} \hbox{Tr}  e^{-\beta {\cal H'}_L} = 
 \lim_{L\rightarrow \infty}\lim_{N\rightarrow \infty}
  \hbox{ Tr } [{\cal T}(u) \overline{\cal T}(u)]^{N/2} |_{u=\beta/N}
\end{equation}
where ${\cal H'}_L$ differs from ${\cal H}_L$ by the sublattice gauge 
transformation,
$ c_{n,\sigma} \rightarrow (-1)^n c_{n,\sigma}$ which does not affect
thermodynamic behaviors.
We regard the resulting system as a fictitious two-dimensional model 
on a $L\times N$ square lattice,
where  $N$ is the extension in the fictitious (imaginary time) direction, 
sometimes referred to as the Trotter number. 
Now by looking at the system in a 90$^\circ$ rotated frame, 
it is natural to define
the ``quantum transfer matrix'' (QTM) by 
\begin{equation}
{\cal T}_{{\rm QTM}}(u,v) = \bigotimes^{N/2} R(-u,v)  
\otimes \widetilde{R}(v,u).
\end{equation}
The interchangeability of the
two limits ($ L, N \rightarrow \infty$) leads to the following expression,
\begin{equation}
Z= \lim_{N\rightarrow \infty} \lim_{L\rightarrow \infty}
  \hbox{ Tr } \left[{\cal T}_{{\rm QTM}}\left(u=\frac{\beta}{N},0\right) 
\right]^{L}.
\end{equation}
There is a gap between the largest and the second largest eigenvalues
of ${\cal T}_{{\rm QTM}}(u,0)$ for finite $\beta$.
Therefore, 
we have a formula for the free energy per site,
\begin{equation}
f= -k_B T \lim_{N\rightarrow \infty}\ln\Lambda_{{\rm max}}
                   \left(u=\frac{\beta}{N},0\right).
\end{equation}
where $\Lambda_{{\rm max}}(u,0)$ denotes 
the largest eigenvalue of ${\cal T}_{{\rm QTM}}(u,0)$.
Now the evaluation of the free energy reduces to that of the single eigenvalue
  $\Lambda_{{\rm max}}$. 
Of course, a sophisticated treatment is necessary in taking the 
Trotter limit $N\rightarrow \infty$ as $u$ now explicitly depends on it.
The following sections are devoted to this analysis.

\begin{figure}[tb]
  \begin{center}
    \leavevmode
    \includegraphics[width=0.45\textwidth]{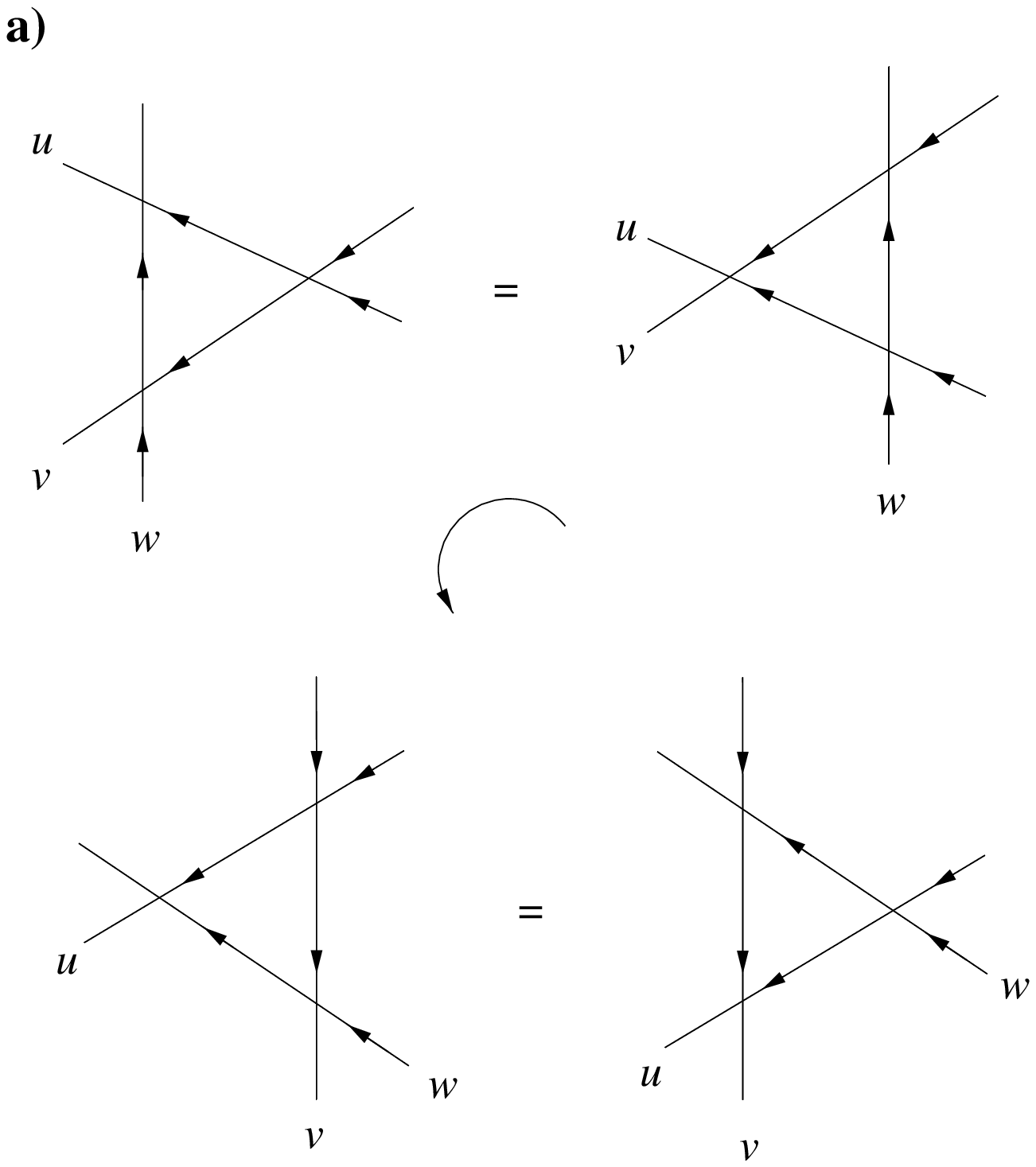}
    \includegraphics[width=0.5\textwidth]{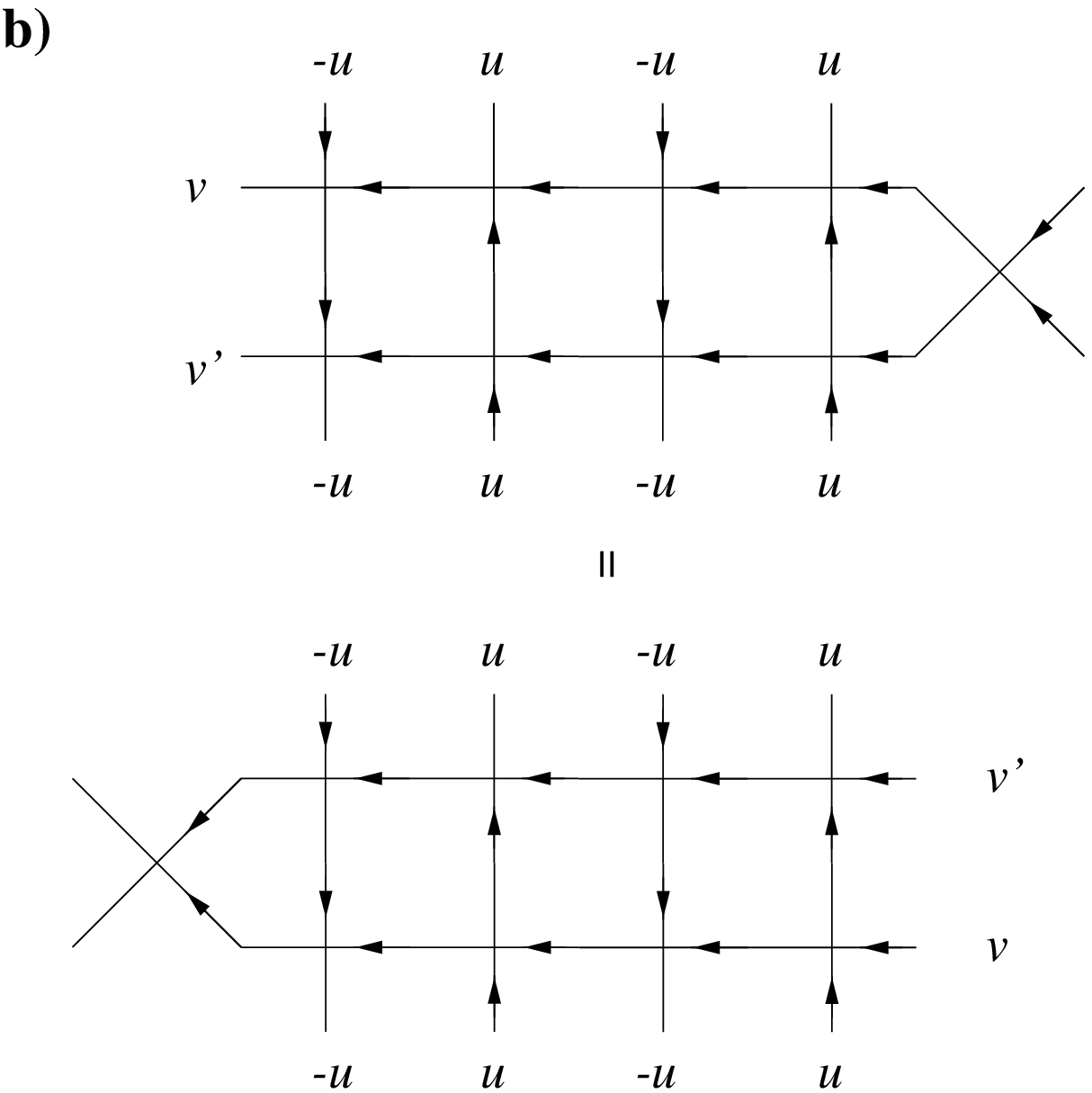}
    \caption{a) Graphical depiction of the fundamental Yang-Baxter equation
for $R$ and the associated one for $R$ and $\widetilde R$ obtained
through rotation. b) ``Railroad proof'' for the commutation of two
QTM's with any spectral parameters $v$ and $v'$. Due to a) the intertwiner
for $R$ vertices is identical to the intertwiner for $\widetilde R$ 
vertices.}
    \label{fig:fig1ab}
  \end{center}
\end{figure}

A general comment is in order.
It seems somewhat redundant to define  ${\cal T}_{{\rm QTM}}(u,v)$ as
we only need the value at $v=0$.
This formulation, however, manifests the integrability structure and
the existence of infinitely many conserved quantities.
This is best seen in the commutativity of transfer matrices
\begin{equation}
[{\cal T}_{{\rm QTM}}(u,v), {\cal T}_{{\rm QTM}}(u,v')]=0,
\end{equation}
with fixed $u$. One can prove this by showing
that two QTMs are intertwined by the same
$R$ operator as for the row-to-row case.
The outline of the proof is graphically 
demonstrated in Fig.\ref{fig:fig1ab}.
The existence of the parameter labeling the family of commuting matrices
makes the subsequent analysis much more transparent.

\section{Diagonalization of the Quantum Transfer Matrix}
It has been an issue of current interest to find 
an explicit algorithm of the 
diagonalization of the row-to-row transfer matrix
of Shastry's model or its fermion version.
The standard tool in such studies, the  quantum inverse scattering
method, has been applied and turned out to be successful
after elaborate calculations \cite{RamMar96}. We also refer to
the analytic Bethe ansatz study \cite{YueDeg97}.

Here we want to diagonalize the QTM.
At first glance, the diagonalization scheme for this looks
quite different from the row-to-row case.
The QTM has a complicated inhomogeneous structure
seemingly demanding much more effort.
Fortunately, this is not true. 
The crucial observation is, as remarked in the
previous section, that QTMs share
the same intertwining operator with
the row-to-row transfer matrices.
In view of QISM, this results into identical
operator algebras which allows for the diagonalization
of the trace of the monodromy matrix.
(Note that we adopt periodic boundaries in the Trotter
direction.)
Thus, the  eigenvalue equation of the QTM
involves the same combinations of ``dress functions''
(in the terminology of the analytic Bethe ansatz)
as in the row-to-row case.
One only has to replace the vacuum expectation values taking account of
the quantum space and the inhomogeneity.

We define state vectors $|i\>, i=1,\cdots, 4 $ by
\begin{equation}
|1\> =|+,-\> , \quad |2\>=|+,+\>, \quad |3\>=|-,-\>, \quad |4\>=|-, +\>.
\end{equation}
A convenient vacuum in the present study may be 
$|\Omega\> := |1,4,1,4,\cdots,1,4\>$.
Then the vacuum expectation values read
\begin{equation}
A_i = \<\Omega| {\cal T}_{i,i}|\Omega\> 
    = (R^{i,1}_{1,i}(-u,v) R^{4,i}_{i,4}(v,u))^{N/2}  \quad i=1, \cdots, 4.
\end{equation}
Substituting explicit elements for $R$, we find the relations
\begin{eqnarray*}
A_1/A_2 &=&   \left ( \frac{(1-z_-(w) z_+(x))(1-z_+(w)z_+(x))}
                  {(1+z_-(w) z_+(x))(1+z_+(w)z_+(x))} \right )^{N/2} \\
A_2    &=& A_3  \\
A_4/A_2 &=& 
           \left ( \frac{(1+z_-(w)/ z_-(x))(1+z_+(w)/z_-(x))}
                     {(1-z_-(w)/ z_-(x))(1-z_+(w)/z_-(x))}
                                             \right )^{N/2} \\
A_2 &=&  \left(
         \frac{\cos^2(u-v)\cos^2(u+v)}{(1+e^{4x})(1+e^{4w})}
         e^{2h(w)}(\frac{1}{z_-(w)}-\frac{1}{z_-(x)})
                    (z_+(x)+\frac{1}{z_-(w)}) \right )^{N/2}
\end{eqnarray*}
where we have introduced the parameterizations $x$, $w$ for $v$, $u$,
\begin{equation}
e^{2x} = \tan v,\qquad e^{2w} = \tan u,
\end{equation}
and the functions
\begin{equation}
z_{\pm}(x) = e^{2h(x)\pm 2x}, \quad 2h (x) = 
-\sinh^{-1}\left(\frac{U}{4\cosh 2x}\right).
\end{equation}
Now that we have the explicit vacuum expectation values, 
the eigenvalue can be written down directly thanks to the above argument
%
%
\begin{equation}
\begin{split}
\frac{\Lambda(v)} {A_2}&=
    e^{\beta (\mu+H/2)} \frac{A_1}{A_2}
    \prod_{j=1}^{m} e^{2x} \frac{1+z_j z_-(x)}{1-z_j z_+(x)}\\
 &+ e^{2\beta\mu}  \prod_{j=1}^{m} -e^{2x} \frac{1+z_j z_-(x)}{1-z_j z_+(x)}
     \prod_{\alpha=1}^{\ell} -\frac{z_-(x)-1/z_-(x)-2iw_{\alpha}+3 U/2}
                         {z_-(x)-1/z_-(x)-2iw_{\alpha}+ U/2} \\
 &+ \prod_{j=1}^{m} -e^{-2x}\frac{1+z_+(x)/z_j}{1-z_-(x)/z_j}
     \prod_{\alpha=1}^{\ell}-\frac{z_-(x)-1/z_-(x)-2iw_{\alpha}- U/2}
                         {z_-(x)-1/z_-(x)-2iw_{\alpha}+ U/2} \\
 &+ e^{\beta(\mu-H/2)} \frac{A_4}{A_2}
     \prod_{j=1}^{m} e^{-2x}\frac{1+z_+(x)/z_j}{1-z_-(x)/z_j}.
\label{eigenval}         
\end{split}
\end{equation}
Note  that we imposed a non-vanishing chemical potential and magnetic field
at the last stage, as they merely
lead to trivial modifications in $\Lambda$ due to twisted boundary
conditions for the QTM \cite{JutKluSuz97}.

The parameters $\{z_j\}, \{w_{\alpha}\}$ satisfy the BAE,
\begin{equation}
\begin{split}
e^{\beta(\mu-H/2)}
\left(  \frac{(1+z_-(w)/z_j)(1+z_+(w)/z_j)}{(1-z_-(w)/z_j)(1-z_+(w)/z_j)} 
   \right)^{N/2} 
   &= -(-1)^m \prod_{\alpha=1}^{\ell} -
      \frac{z_j-1/z_j-2iw_{\alpha}-U/2}{z_j-1/z_j-2iw_{\alpha}+U/2}, \\
e^{2\beta\mu} \prod_{j=1}^{m}
    \frac{z_j-1/z_j-2iw_{\alpha}+U/2}{z_j-1/z_j-2iw_{\alpha}-U/2}
  &= -\prod_{\beta=1}^{\ell}
    \frac{2i (w_{\alpha}-w_{\beta})-U}{2i (w_{\alpha}-w_{\beta})+U}.
\label{bae-eq}
\end{split}
\end{equation}
Here some remarks are in order:
\begin{enumerate}
\item
Although the validity of expression (\ref{eigenval})
is a logical consequence, it would be a good exercise to verify this
form for one-particle states.
For example, we take ${\cal T}_{3,4}(\nu)|\Omega\>$ as a
representative, and calculate its eigenvalue.
A standard argument in QISM leads to the following ``wanted terms''.
\begin{equation}
\begin{split}
  \hbox{wanted terms }&= \Biggl[ A_1
  \frac{R^{3,1}_{1,3}(v,\nu)}{R^{4,1}_{1,4}(v,\nu)} + A_2
  \Biggl(\frac{R^{3,2}_{2,3}(v,\nu)}{R^{4,2}_{2,4}(v,\nu)}-
  \frac{R^{3,2}_{1,4}(v,\nu)R^{4,1}_{2,3}(v,\nu)}
  {R^{4,2}_{2,4}(v,\nu)R^{4,1}_{1,4}(v,\nu)}\Biggr) \\ &+ A_3
  \frac{R^{3,3}_{3,3}(v,\nu)}{R^{4,3}_{3,4}(v,\nu)} + A_4
  \frac{R^{4,4}_{4,4}(\nu,v)}{R^{4,3}_{3,4}(\nu,v)} \Biggr] {\cal
    T}_{3,4}(\nu)|\Omega\>
\label{one-particle}
\end{split}
\end{equation}
A straightforward however lengthy calculation shows the coefficient in
(\ref{one-particle}) is equal to (\ref{eigenval}) with $m=1$, $\ell=0$,
$z_1=z_-(1/2\log(\tan\nu))$.
\item
We have verified that (\ref{eigenval}) gives 
the largest eigenvalue identical to the one obtained by brute force
diagonalizations of finite systems up to $N=6$.  The groundstate lies
in the sector $m=N$, $\ell=N/2$.  For the repulsive case and $\mu=H=0$,
$z_j$'s are all on the imaginary axis, while $w_{\alpha}$'s are real.
\item
The free-fermion partition function is 
easily recovered for $U=0$.
\item
Starting from another vacuum $|\Omega'\>=|2,3,\cdots\>$, 
one reaches a different expression.  The resultant one is actually
equivalent after negating $U$ and exchanging $H/2 \leftrightarrow
\mu$, namely, after the partial particle-hole transformation.
\end{enumerate}

\section{Associated auxiliary problem of difference type}

The thermodynamics leading to the free energy is encoded in the solution to
the BAE (\ref{bae-eq}) in the limit $N\to\infty$. For finite $N$ it is possible
to solve the BAE numerically. However, for large $N$ it is quite complicated to
find the numerical solution even for the ground state.
Furthermore,
in the Trotter limit $N\to\infty$ the
roots $\{v_k,w_k\}$ accumulate at infinity. This is similar to other models
(Heisenberg model, $t-J$ model) where the solutions of BAE
of the QTM tend to the origin 
\cite{Klu92,Klu93,DesVeg92,JutKlu96,JutKluSuz97,JutKluSuz97b}. 
It represents the main problem in analyzing the limit $N\to\infty$
directly on the basis of the BAE.  
To overcome this difficulty one can express the solution of the BAE 
by a system of non-linear integral equations. This has been done for several
models 
\cite{KlumB90,KlumBP91,Klu92,Klu93,DesVeg92,JutKlu96,JutKluSuz97,JutKluSuz97b}. 

The first problem to be overcome is the involved structure of BAE
(\ref{bae-eq}). Upon introducing the quantities
\begin{equation}
s_j =\frac{1}{2i} \left(z_j-\frac{1}{z_j}\right),
\end{equation}
the equations (\ref{bae-eq}) take a difference form in the rapidities
$\{s_j\}, \{w_{\alpha}\}$:
\begin{align}
{\rm e}^{-\beta(\mu-H/2)}\phi(s_j) &= 
-\frac{q_2(s_j-i\gamma)}{q_2(s_j+i\gamma)},\\
{\rm e}^{-2\beta\mu}\frac{q_2(w_\alpha+2i\gamma)}{q_2(w_\alpha-2i\gamma)}&=-
\frac{q_1(w_\alpha+i\gamma)}{q_1(w_\alpha-i\gamma)},\label{aux-bae}
\end{align}
with
\begin{align}
  q_1(s)&=\prod_j(s-s_j), \quad q_2(s)=\prod_\alpha(s-w_\alpha),\quad \gamma=
\frac{U}{4},\\
\intertext{and}
  \phi(s)&=
\left(  \frac{(1-z_-(w)/z(s))(1-z_+(w)/z(s))}{(1+z_-(w)/z(s))(1+z_+(w)/z(s))} 
   \right)^{N/2},\\
z(s) &= i s(1+\sqrt{(1-1/s^2)}).
\end{align}
The function $z(s)$ possesses two branches. The standard (``first'') branch 
is chosen by the requirement $z(s)\simeq 2is$ for large values of $s$, and
the branch cut line $[-1,1]$ (corresponding to a cut for $\sqrt{z}$ 
from $-\infty$ to $0$.) We will not refer explicitly
to the second branch of $z(s)$ in
this work. However, for $\phi(s)$ both branches will be used. In order to
distinguish between the first and second one we will use the notation
$\phi^+(s)$ and $\phi^-(s)$, respectively. 

\begin{figure}[tb]
  \begin{center}
    \leavevmode
    \includegraphics[width=0.45\textwidth]{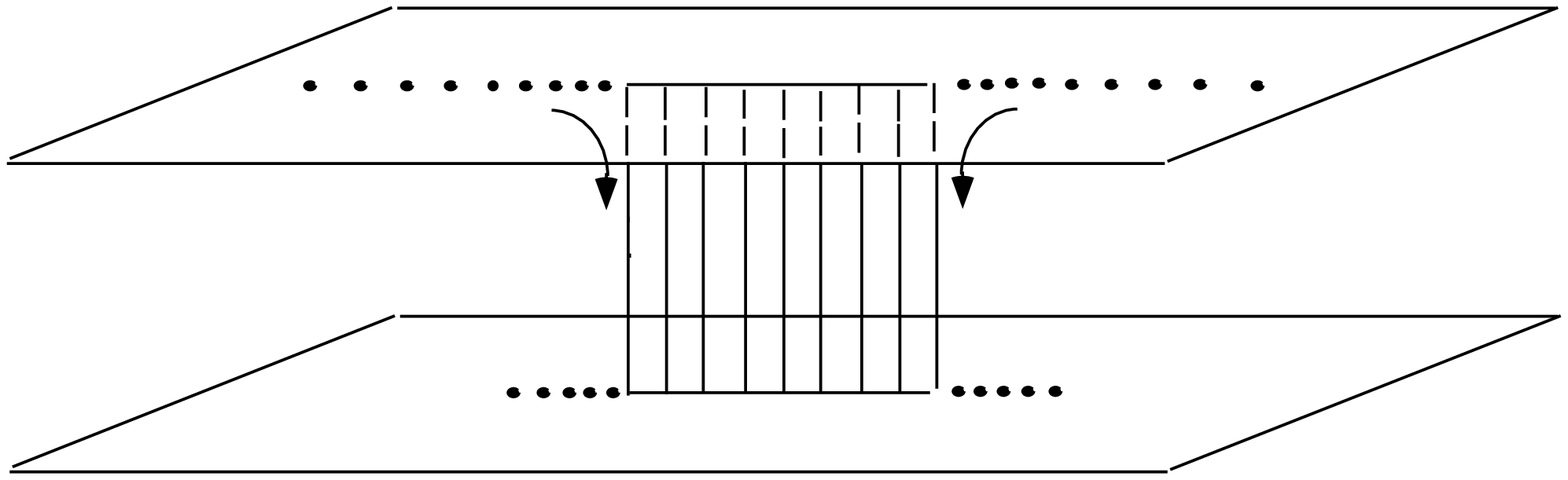}
    \caption{Depiction of the flow of rapidities $s_j$ from the first 
(upper) Riemann sheet to the second (lower) one for decreasing temperature.}
    \label{fig:fig2}
  \end{center}
\end{figure}

In Fig.\ref{fig:fig2} the distribution of
rapidities $s_j$ is shown for $N\to\infty$. Note that there are infinitely 
many 
rapidities on the first (upper) sheet and finitely many on the second (lower)
sheet. The number of rapidities on the second sheet is increasing with
decreasing temperature thus resulting into a flow from the first to the
second sheet.

Note that the general expression (\ref{eigenval}) is quite complicated, but
simplifies considerably at $v=0$ and $u \rightarrow 0$
\begin{equation}
\Lambda(v=0) = e^{\beta \gamma}
          (1+e^{\beta (\mu+H/2)} )(1+e^{\beta (\mu-H/2)} )
          u^N \prod_{j=1}^{m} z_j .
\label{eig_prac}
\end{equation}

Lastly, we want to comment on the difference type property of (\ref{aux-bae}).
These equations are ``BAE compatible'' with the
following eigenvalue equation of an ``auxiliary transfer matrix'' 
\begin{equation}
  \Lambda^{\rm aux}(s) = 
   \lambda_1(s)+\lambda_2(s)+\lambda_3(s)+\lambda_4(s),
\end{equation}
\begin{equation}
  \begin{aligned}
    \lambda_1(s) &={\rm e}^{\beta(\mu+H/2)}
      \frac{\phi(s-i\gamma)}{q_1(s-i\gamma)}\,,\quad & 
    \lambda_2(s) &={\rm e}^{2\beta\mu}
      \frac{q_2(s-2i\gamma)}{q_2(s)q_1(s-i\gamma)}\,,\\
    \lambda_3(s) &=\frac{q_2(s+2i\gamma)}{q_2(s)q_1(s+i\gamma)},&
    \lambda_4(s) &={\rm e}^{\beta(\mu-H/2)}
      \frac{1}{\phi(s+i\gamma)\,q_1(s+i\gamma)}\,.
  \end{aligned}
\label{eig-aux}
\end{equation}
The reason is obvious, $\Lambda^{\rm aux}(s)$ is pole free under 
BAE (\ref{aux-bae}) just like the original QTM.
The construction (\ref{eig-aux}) at this point is purely mathematical,
however it will be the starting point of the derivation of integral 
equations in the next section.

\section{Non-linear integral equations}
\label{sec:non-linear-integral-equations}
In this section we are concerned with the derivation of well posed
integral equations equivalent to the nested BAE
for the largest eigenvalue of the QTM. 
(We restrict ourselves to the case $U>0$ and point out that $U<0$ is
simply obtained via a particle-hole transformation.)
We introduce a 
set of auxiliary functions which satisfy a set of closed functional
equations which later on are transformed into integral form. The 
explicit expression of the functions $\b$, $\bb$, $\c$, $\bc$
which proved useful is
\begin{equation}
  \begin{split}
    \b&=
    \frac{\bl{1}+\bl{2}+\bl{3}+\bl{4}}{l_1+l_2+l_3+l_4},\qquad
    \bb=
    \frac{l_1+l_2+l_3+l_4}{\bl{1}+\bl{2}+\bl{3}+\bl{4}},\\
    \c&=\frac{l_1+l_2}{l_3+l_4}\cdot
    \frac{\bl{1}+\bl{2}+\bl{3}+\bl{4}}
         {l_1+l_2+l_3+l_4+\bl{1}+\bl{2}+\bl{3}+\bl{4}},\\
    \bc&=\frac{\bl{3}+\bl{4}}{\bl{1}+\bl{2}}\cdot
    \frac{l_{1}+l_{2}+l_{3}+l_{4}}
         {l_1+l_2+l_3+l_4+\bl{1}+\bl{2}+\bl{3}+\bl{4}},
  \end{split}
\label{auxFunct}
\end{equation}
where the functions $l_j$ and $\bl{j}$ are closely related to the $\lambda_j$
defined in (\ref{eig-aux})
\begin{equation}
  \begin{split}
l_j(s)&=\lambda_j(s-i\gamma)\cdot\e^{\beta H}{\phi^+(s)\phi^-(s)},\\
\bl{j}(s)&=\lambda_j(s+i\gamma).
  \end{split}
\end{equation}

The main observation in connection with the functions defined in 
(\ref{auxFunct}) is based on elementary facts of the theory of complex
functions. In particular any analytic function on the complex plane
is entirely determined by its singularities, i.e. poles and branch cuts,
as well as its asymptotic behavior at infinity. Below we will show
that the singularities of $\ln\b$, $\ln\c$ etc. on the {\it entire complex}
plane are exhausted by the singularities of $\ln(1+\b)$, $\ln(1+\c)$ etc.
close to the {\it real axis}\footnote{The relevant singularities are 
distributed exactly on the real axis for vanishing external fields. 
For this case the subsequent treatment can be taken literally. For finite
external fields $h$, $\mu$ deviations from the real axis occur. The following
reasoning still applies {\it mutatis mutandis}.}. Furthermore all the
involved functions show constant asymptotics for $N$ finite. 
Hence there exists a suitable
integral representation of $\ln\b$, $\ln\c$ etc. in terms of
$\ln(1+\b)$, $\ln(1+\c)$ etc. The latter functions will be abbreviated
by
\begin{equation}
  \begin{split}
    \B&=1+\b=
    \frac{l_1+l_2+l_3+l_4+\bl{1}+\bl{2}+\bl{3}+\bl{4}}{l_1+l_2+l_3+l_4},\\
    \bB&=1+\bb=
    \frac{l_1+l_2+l_3+l_4+\bl{1}+\bl{2}+\bl{3}+\bl{4}}
    {\bl{1}+\bl{2}+\bl{3}+\bl{4}},\\
    \C&=1+\c=\frac{l_1+l_2+l_3+l_4}{l_3+l_4}\cdot
    \frac{l_3+l_4+\bl{1}+\bl{2}+\bl{3}+\bl{4}}
         {l_1+l_2+l_3+l_4+\bl{1}+\bl{2}+\bl{3}+\bl{4}},\\
    \bC&=1+\bc=\frac{\bl{1}+\bl{2}+\bl{3}+\bl{4}}{\bl{1}+\bl{2}}\cdot
    \frac{l_{1}+l_{2}+l_{3}+l_{4}+\bl{1}+\bl{2}}
         {l_1+l_2+l_3+l_4+\bl{1}+\bl{2}+\bl{3}+\bl{4}}.
  \end{split}
\label{auxFunct3}
\end{equation}
Quite generally all the above auxiliary functions have a product
representation with factors of the type $...+l_3+l_4+\bl{1}+\bl{2}+...$ .
As a matter of the BAE the poles of each $l_j$ and
$\bl{j}$ function in $...+l_3+l_4+\bl{1}+\bl{2}+...$ are canceled by
the neighboring terms. Poles can only ``survive'' if such a string
does not begin with $l_1$ or does not end with $\bl{4}$. There are
extended singularities (cuts) due to the function $\phi$ appearing 
in the definition of $\lambda_1$ and $\lambda_4$. Hence all terms
$l_1+l_2+...$ and $...+\bl{3}+\bl{4}$ possess branch cuts along
$[-1,1]+2i\gamma$ and $[-1,1]-2i\gamma$, respectively. Furthermore,
terms like $...+l_3+l_4$ and $\bl{1}+\bl{2}+...$ have branch cuts along
$[-1,1]$. However in combinations $...+l_4+\bl{1}+...$ the branch cut
due to the $\phi$ function disappears, because
\begin{equation}
l_4(s)+\bl{1}(s)=\e^{\beta(\mu+H/2)}\frac{\phi^+(s)+\phi^-(s)}{q_1(s)},
\end{equation}
and $\phi^+(s)+\phi^-(s)$ is analytic everywhere.

Inspecting the function $\lambda_1+\lambda_2+\lambda_3+\lambda_4$
more closely we find poles of order $N/2$ at $s_0-i\gamma$ and $i\gamma-s_0$
where
\begin{equation}
z(s_0)=z_-(w),\qquad 2is_0\simeq N/\beta \hbox{ for large } N.
\label{bare}
\end{equation}
In addition we find zeros and branch cuts on the lines $\Im(s)=\pm\gamma$
which we denote by
\begin{equation}
  \begin{split}
    \ln[\lambda_1(s)+\lambda_2(s)+\lambda_3(s)+\lambda_4(s)]
    \sing&-\frac{N}{2}\ln[(s+i\gamma-s_0)(s+s_0-i\gamma)]\\
    &+L_-(s+i\gamma)+L_+(s-i\gamma),\\
  \end{split}
\label{DLam}
\end{equation}
where $\sing$ denotes that left and right hand sides have the same
singularities on the entire plane, and
$L_\pm$ are suitable functions possessing the desired singularities
and being analytic otherwise. (The existence of such functions can be
proved quite easily. An explicit expression is given by contour integrals
of the type (\ref{jump}).) From (\ref{DLam}) we find the following 
singularities
\begin{equation}
  \begin{split}
    \ln[l_1(s)+l_2(s)+l_3(s)+l_4(s)]
\sing&-\frac{N}{2}\ln[(s-s_0)(s+s_0-2i\gamma)]+\ln[{\phi^+(s)\phi^-(s)}]\\
&+L_-(s)+L_+(s-2i\gamma),\\
    \ln[\bl{1}(s)+\bl{2}(s)+\bl{3}(s)+\bl{4}(s)]
    \sing&-\frac{N}{2}\ln[(s+2i\gamma-s_0)(s+s_0)]\\
&+L_-(s+2i\gamma)+L_+(s),
  \end{split}
\label{DL}
\end{equation}
From this, and (\ref{auxFunct},\ref{auxFunct3}) and the identity
\begin{equation}
\phi^+(s)\phi^-(s)=
\left[\frac{(s-s_0)(s+s_0-2i\gamma)}
{(s+s_0)(s-s_0+2i\gamma)}\right]^{N/2},
\end{equation}
we find the singularities
\begin{equation}
  \begin{split}
\ln\b(s)\sing&\ L_-(s+2i\gamma)+L_+(s)-L_-(s)-L_+(s-2i\gamma),\\
\ln\B(s)\sing&\ -L_-(s)+\hbox{rest},\\
\ln\bc(s)-\ln\bC(s)\sing&\ L_-(s)-L_+(s)+\hbox{rest},
  \end{split}
\label{jump2}
\end{equation}
where ``rest'' indicates singularities not located on the real axis.

Next we introduce the notation 
\begin{equation}
(g\circ f)(s)=\int_{\cal L}g(s-t)f(t)dt,
\end{equation}
for the convolution of two functions $g$ and $f$
with contour ${\cal L}$ surrounding the real axis 
at infinitesimal distance above and below in anticlockwise
manner. From Cauchy's theorem we find for any function $f$ analytic
above and below the real axis
\begin{equation}
(k\circ f)(x\pm i\epsilon)=(k\circ f)(x)+f(x\pm i\epsilon),
\hbox{ where }
k(s)=\frac{1}{2\pi i}\frac{1}{s}.
\label{jump}
\end{equation}
For further convenience we introduce the functions
\begin{equation}
  \begin{split}
K_1(s)&=k(s)-k(s+2i\gamma)=\frac{\gamma/\pi}{s(s+2i\gamma)},\cr
{\overline{K_1}}(s)&=-k(s)+k(s-2i\gamma)=\frac{\gamma/\pi}{s(s-2i\gamma)},\\
K_2(s)&=k(s-2i\gamma)-k(s+2i\gamma)=
\frac{2\gamma/\pi}{s^2+4\gamma^2},\cr
  \end{split}
\label{kernels}
\end{equation}
which will play the role of integral kernels.
From (\ref{jump2},\ref{jump},\ref{kernels}) we find
\begin{equation}
[K_2\circ\ln\B+{\overline{K_1}}\circ(\ln\bc-\ln\bC)]
\sing L_-(s+2i\gamma)+L_+(s)-L_-(s)-L_+(s-2i\gamma).
\label{Komb1}
\end{equation}
Upon comparing (\ref{jump2},\ref{Komb1}) we conclude
\begin{equation}
\ln\b(s)=K_2\circ\ln\B+{\overline{K_1}}\circ(\ln\bc-\ln\bC)+\hbox{const},
\end{equation}
as both sides are complex functions with identical singularities.
(The difference function is entire, i.e. analytic on the entire
complex plane. Furthermore the difference function is bounded, hence 
it is constant.) The constant is computed from considering the
asymptotic behavior at $s\to\infty$
\begin{equation}
\hbox{const}=-\beta H.
\end{equation}

For the derivation of the second type of integral equation we
define an intermediate set of auxiliary functions
\begin{equation}
  \begin{split}
    \t&=\frac{l_1+l_2}{l_3+l_4},\qquad 
    \T=1+\t=\frac{l_1+l_2+l_3+l_4}{l_3+l_4},\\
    \bt&=\frac{\bl{3}+\bl{4}}{\bl{1}+\bl{2}},\qquad
    \bT=1+\bt=\frac{\bl{1}+\bl{2}+\bl{3}+\bl{4}}{\bl{1}+\bl{2}}.
  \end{split}
\label{auxFunct4}
\end{equation}
Quite similar to the above derivation we conclude the identity
\begin{equation}
\ln\t(s)=\frac{N}{2}\ln\frac{s+s_0}{s+s_0-2i\gamma}+
\beta(\mu+H/2)+\ln\phi(s)
-{\overline{K_1}}\circ(\ln\bB+\ln\bT).
\label{eqt}
\end{equation}
Next we deform the integration contour for $\ln\bB$ in (\ref{eqt})
from a narrow loop around the real axis to a wide loop consisting
of the two horizontal lines $\Im s=\pm\alpha$, with $0<\alpha\le\gamma$. 
The corresponding
convolution is denoted by ``$\wcirc$''
\begin{equation}
{\overline{K_1}}\circ\ln\bB={\overline{K_1}}\wcirc\ln\bB-\ln\bB,
\end{equation}
and the additional contribution is due to the residue of 
$\overline{K_1}$.
Taking into account of (\ref{auxFunct},\ref{auxFunct4}) we find
\begin{equation}
\ln\c=\ln\t-\ln\bB,\qquad \Delta\ln\bT=\Delta\ln\bC,
\end{equation}
where $\Delta f(x)=f(x+i0)-f(x-i0)$ denotes the discontinuity
along the real axis.
Therefore, (\ref{eqt}) turns into
\begin{equation}
\ln\c(s)=\frac{N}{2}\ln\frac{s+s_0}{s+s_0-2i\gamma}+\beta(\mu+H/2)+\ln\phi(s)
-{\overline{K_1}}\wcirc\ln\bB-{\overline{K_1}}\circ\ln\bC.
\label{eqc}
\end{equation}

Lastly, we perform the limit $N\to\infty$ in the above equations yielding
\begin{equation}
  \begin{split}
\ln\b&=-\beta H+K_2\wcirc\ln\B+{\overline{K_1}}\circ(\ln\bc-\ln\bC),\cr
\ln\c&=-\beta{U}/{2}+\beta(\mu+H/2)+\ln\phi
-{\overline{K_1}}\wcirc\ln\bB-{\overline{K_1}}\circ\ln\bC,\cr
\ln\bc&=-\beta{U}/{2}-\beta(\mu+H/2)-\ln\phi
+K_1\wcirc\ln\B+K_1\circ\ln\C,
  \end{split}
\label{all}
\end{equation}
where the equation for $\ln\bc$ has been derived in analogy to the one
for $\ln\c$, and the function $\phi$ has the simplified expression
\begin{equation}
\ln\phi(x)=-2\beta i x\sqrt{1-1/x^2}.
\end{equation}
We want to point out that the function $\b$ will be evaluated 
on the lines $\Im s=\pm\alpha$. The functions $\c$ and $\bc$ need only
be evaluated on the real axis infinitesimally above and below the interval
$[-1,1]$. Also the convolutions involving the ``$\c$ functions'' in
(\ref{all}) can be restricted to a contour surrounding $[-1,1]$ as
these functions are analytic outside.

The detailed derivation of integral expressions for the largest eigenvalue
$\Lambda$ of the QTM is deferred to Appendix A. Here we restrict ourselves 
to a compilation of the most relevant results
\begin{equation}
  \begin{split}
2\pi i\ln\Lambda=&2\pi i\beta(\mu+U/4)+\int_{\cal L}[\ln z(s)]'
\ln\left(1+\c+\bc\right)ds\\
&-\frac{1}{2}
\int_{\cal L}\left[\ln\frac{z(s-2i\gamma)}{z(s)}\right]'\ln\B(s)ds
-\frac{1}{2}
\int_{\cal L}\left[\ln\frac{z(s+2i\gamma)}{z(s)}\right]'\ln\bB(s)ds,\\
=&2\pi i\beta(\mu+H/2+U/4)+\int_{\cal L}[\ln z(s)]'
\ln\left(1+\c+\bc\right)ds\\
&-\int_{\cal L}\left[\ln\frac{z(s-2i\gamma)}{z(s)}\right]'\ln\B(s)ds,\\
=&-2\pi i\beta\frac{U}{4}+\int_{\cal L}[\ln z(s)]'
\ln\frac{1+\c+\bc}{\bc}ds
+\int_{\cal L}\left[\ln{z(s-2i\gamma)}\right]'\ln\C(s)ds.
  \end{split}
\label{int-eig}
\end{equation}
The last two expressions are of particular importance to our further
numerical and analytical treatment.

Finally, we want to comment on the structure of the equations determining the 
thermodynamical properties of the Hubbard model. In contrast to 
long-range interaction systems \cite{KatKur96,GebGirRuc94}
we have to solve a set of subsidiary equations (\ref{all}) for the 
``distribution functions'' $\b, \c,$ and $\bc$ before
evaluating the free energy (\ref{int-eig}). Obviously, the dynamics of
the elementary excitations of the nearest-neighbor systems is more
involved than those of \cite{KatKur96,GebGirRuc94} which may be viewed
as ``free particles with exclusion statistics''.

\section{Numerical Results}
For the numerical treatment of equations (\ref{all},\ref{int-eig})
we rewrite them in terms of usual convolutions of functions of a real 
variable
\begin{equation}K\ast f=\int_{-\infty}^\infty K(x-y)\, f(y)\,{\rm d}y.\end{equation}
For the functions (\ref{auxFunct}) evaluated on the contours involved
in (\ref{all},\ref{int-eig}) we use the notation
${\frak b}^\pm$, ${\frak c}^\pm$
and $\overline{{\frak c}^\pm}$
\begin{equation}
  {\frak b}^\pm(x)={\frak b}(x\pm i\alpha),\quad
  {\frak c}^\pm(x)={\frak c}(x\pm i\,0),\quad
  \overline{{\frak c}^\pm}(x)=\overline{{\frak c}}(x\pm i\,0),
\end{equation}
where the shift $\alpha$ in ${\frak b}$ is arbitrary but fixed with
\mbox{$0<\alpha\le\gamma$}. (For many numerical calculation we take
$\alpha=2\gamma/3$ and especially $\alpha=\gamma$.) 
Furthermore, we introduce the following relations:
\begin{align}
    {\frak B}^\pm&:=1+{\frak b}^\pm,
    &\overline{{\frak B}^\pm}&:=1+1/{\frak b}^\pm,\nonumber\\ 
    {\frak C}^\pm&:=1+{\frak c}^\pm,
    &\overline{{\frak C}^\pm}&:=1+\overline{{\frak c}^\pm},
    \label{BC-relations}\\ 
    \Delta\log{{\frak C}}&:=\log({\frak C}^+/{\frak C}^-),
    &\Delta\log{\overline{{\frak C}}}&:=
    \log(\overline{{\frak C}^+}/\overline{{\frak C}^-}).\nonumber
\end{align}
Thus (\ref{all}) is written in the form
\begin{equation}
  \begin{split}
  \log{\frak b}^{\pm}&=
  -\beta H
  -K_{2,\pm\alpha-\alpha}\ast\log{\frak B}^{+}
  +K_{2,\pm\alpha+\alpha}\ast\log{\frak B}^{-}
  -\overline{K}_{1,\pm\alpha}\ast\Delta
         \log(\overline{{\frak c}}/\overline{{\frak C}}),\\
  \log{\frak c}^\pm&= 
  \Psi_{\frak c}^\pm
  +\overline{K}_{1,-\alpha}\ast\log\overline{{\frak B}^+}
  -\overline{K}_{1,\alpha}\ast\log\overline{{\frak B}^-}
  +\overline{K}_{1,0}\ast\Delta\log\overline{{\frak C}}
  \pm{\textstyle\frac{1}{2}}\Delta\log\overline{{\frak C}},\\
  \log\overline{{\frak c}^\pm}&= 
  \overline{\Psi}_{\frak c}^\pm
  -{K}_{1,-\alpha}\ast\log{{\frak B}^+}
  +{K}_{1,\alpha}\ast\log{{\frak B}^-}
  -{K}_{1,0}\ast\Delta\log{{\frak C}}
  \pm{\textstyle\frac{1}{2}}\Delta\log{{\frak C}},\\
  \end{split}
\label{nle2-app}
\end{equation}
where
\begin{align}
    \Psi_{\frak c}^\pm&=
      -\beta U/2+\beta(\mu+H/2)+\log\phi_{\pm 0},\\
    \overline{\Psi}_{\frak c}^\pm&=
      -\beta U/2-\beta(\mu+H/2)-\log\phi_{\pm 0},
\end{align}
and we have used the notation $f_\alpha$ for a function $f$ 
with shift of the argument by $i\alpha$
\begin{equation*}
f_\alpha(x)=f(x+i\alpha).
\end{equation*}
In particular $\phi_{\pm 0}$ denotes the function $\phi$ evaluated on the
real axis from above/below.
Notice that the convolution over the terms 
$\Delta\log{{\frak C}}$
and $\Delta\log{\overline{{\frak C}}}$ are determined by Cauchy's
principal value. Remember that these functions vanish outside the interval
$[-1,1]$.

Similarly, from (\ref{int-eig}) we obtain two different relations 
for the eigenvalue 
\begin{align}
  \log\Lambda=
    &-\int_{-1}^{1}{\cal K}\,
           \log[(1+{\frak c}^++\overline{{\frak c}^+})
               (1+{\frak c}^-+\overline{{\frak c}^-})]
    \,{\rm d}x\,\nonumber\\ 
    &+\int_{-\infty}^{\infty}[
        ({\cal K}_{\alpha-2\gamma}-{\cal K}_{\alpha})\,
        \log{\frak B}^+\,-\,
        ({\cal K}_{-\alpha-2\gamma}-{\cal K}_{-\alpha})\,
        \log{\frak B}^-]
  \,{\rm d}x\, +\beta(\mu+H/2+U/4). \nonumber\\
   = &-\int_{-1}^{1}{\cal K}\,
           \log[(1+{\frak c}^++\overline{{\frak c}^+})
               (1+{\frak c}^-+\overline{{\frak c}^-})/
               (\overline{{\frak c}^+}\,\overline{{\frak c}^-})]\,
           {\rm d}x\,\nonumber\\
    &-\int_{-1}^{1}{\cal K}_{-2\gamma}\,
           \log[(1+{\frak c}^+)/(1+{\frak c}^-)]\,{\rm d}x\,\,
    -\beta U/4,
\label{nle-eval2-app}
\end{align}
with
\begin{align}
  {\cal K}(x)=&-\left(2\pi\sqrt{1-x^2}\right)^{-1}
=\left(2\pi i\,x \sqrt{1-1/x^2}\right)^{-1},
\end{align}
and ${\cal K}_\alpha$ is the related
function with shifted argument. The branch of
${\cal K}$ is fixed by the requirement ${\cal K}(x)\simeq 1/(2\pi i x)$
for large $x$.
By means of the relation
\begin{equation*}
  \Delta\log\overline{{\frak c}}=-\Delta\log\phi+\Delta\log{\frak C},
\end{equation*}
the first equation of (\ref{nle2-app}) turns into
\begin{equation}
  \log{\frak b}^{\pm}=
  \Psi_{\frak b}^\pm
  -K_{2,\pm\alpha-\alpha}\ast\log{\frak B}^{+}
  +K_{2,\pm\alpha+\alpha}\ast\log{\frak B}^{-}
  -\overline{K}_{1,\pm\alpha}\ast\Delta
         \log({\frak C}/\overline{{\frak C}}),
\label{nle}
\end{equation}
where
\begin{equation}
    \Psi_{\frak b}^\pm=-\beta U-\beta H
+\log\phi_{\pm\alpha}-\log\phi_{\pm\alpha-2\gamma}.
\end{equation}

For the sake of completeness rather than for further applications
we mention the results for finite Trotter number $N$. All equations above
hold true after the replacement of the ``driving functions'' $\psi$ by
\begin{align*}
  \Psi_{\frak b}^\pm&=-\beta H+\log\phi_{\pm\alpha}
      -\log\phi_{\pm\alpha-2\gamma}
      -{\textstyle\frac{N}{2}\log\frac{x\pm i\alpha-s_0+2i\gamma}
                           {x\pm i\alpha-s_0-2i\gamma}}
      ,\\
  \Psi_{\frak c}^\pm&=+\beta(\mu+H/2)+\log\phi_{\pm 0}
      +{\textstyle\frac{N}{2}\log\frac{x+s_0}
                           {x+s_0-2i\gamma}},\\
  \overline{\Psi}_{\frak c}^\pm&=-\beta(\mu+H/2)-\log\phi_{\pm 0}
      +{\textstyle\frac{N}{2}\log\frac{x-s_0}
                             {x-s_0+2i\gamma}},  
\end{align*}
where $s_0$ is defined in (\ref{bare}). These relations for finite Trotter
number $N$ have been used for a comparison of the results of the integral 
equations with a direct treatment based on the BAE of Sections 3 and 4.
Thus it was possible to ensure the accuracy $(10^{-6})$ of our numerics
based on iterations and fast Fourier transform.

Next we present our numerical results for various physical quantities
and discuss them in terms of the elementary spin and charge
excitations, i.e. ``spinons'' and ``holons'' (plus gapped excitations
based on ``doubly occupied sites'').
Note that at half-filling the system
possesses a charge gap such that the holons do not contribute 
at low temperatures. Furthermore, the hopping integral of the kinetic
energy has been set to $t=1$.

In Fig.\ref{fig:hubU-C} the temperature dependence of 
the specific heat is shown for
densities $n=1$, $0.8$, and $0.5$. For half-filling ($n=1.0$) the specific
heat shows one pronounced temperature maximum for lower values of the
interaction $U$. For stronger $U$ this maximum splits into a 
lower and a higher temperature
maximum which are due to spin and (gapped) charge excitations, respectively.
(These findings agree largely with those of \cite{Shi72a}.)
The picture remains qualitatively true for small dopings ($n=0.8$),
however now the lower temperature peak receives contributions by gapless
charge excitations, hence some weight is shifted from higher 
to lower temperatures. The situation changes quite drastically for 
fillings $n\approx 0.5$.
Here a pronounced maximum in the specific heat is located at a temperature
of about \mbox{$T\approx 0.6$} which
seems rather insensitive to the interaction. 
This is explained by the irrelevance of the onsite interaction
at sufficiently large temperatures, because of the low particle density. 
In addition, we find a maximum at very
low temperatures which depends very sensitively on $U$ 
as well as on the particle density~$n$. In order to clarify the origin
of this additional structure the variation of the specific heat with
$n$ is shown in Fig.\ref{fig:hubConstU-C} for $U=8$. Decreasing the particle 
density $n$ from half-filling
($n=1$) to lower values ($n\approx 0.8$) the ``spin'' maximum at 
lower temperature increases. This picture is changed drastically
below $n\lesssim{}0.8$. Here the ``spin'' maximum and its location 
are suppressed for lower $n$ and a shoulder at a higher temperature
develops into a clear maximum. This new structure in the
specific heat is located at about $T\approx 0.6$ and quite independent 
of $U$ as already mentioned. We interpret this maximum to be
of ``charge'' type.
The complex behavior at intermediate densities $0.5\lesssim{}n\lesssim{}0.7$
is due to a crossover of the ``spin'' and ``charge'' maxima, see also 
Fig.\ref{fig:hubSpinCharge}. For densities $n\approx 1$ the ``spin'' 
maximum is located at finite temperature with finite
height whereas the ``charge'' maximum is located at very small temperature
with small height. For densities close to $n\approx 0$ the situation is 
reversed.

In Fig.\ref{fig:hubU-X} and Fig.\ref{fig:hubConstU-X} 
the magnetic susceptibility $\chi$ 
is presented. Again we begin our
discussion with the half-filled case which is known to correspond to
the Heisenberg spin chain with interaction strength of order $O(t^2/U)$.
In fact, we observe a Heisenberg-like temperature dependence of the
susceptibility with $\chi_{\rm max}$ and $T_{\rm max}$ scaling with $U$ 
and $1/U$ in the range of $U=4,...,8$. Upon doping this behavior remains
qualitatively and quantitatively unchanged even for $n=0.5$. 
Quite generally, the location $T_{\rm max}$ is shifted to lower temperatures,
see Fig.\ref{fig:hubConstU-X}.
The maximal value $\chi_{\rm max}$ decreases for decreasing particle 
density from $n=1$ to $n\approx 0.8$. Below the value
$n\lesssim{}0.8$ the maximum $\chi_{\rm max}$
increases for further lowering of the particle density.
This behavior is qualitatively explained by partially filled bands of 
charge carriers with spin.

In contrast to $\chi$ the charge susceptibility $\kappa$ 
(=$\partial n/\partial \mu$, i.e. compressibility) 
shows a more interesting dependence on the particle density $n$, 
see Fig.\ref{fig:hubU-K} and Fig.\ref{fig:hubConstU-K}. 
At half-filling $\kappa$ shows the expected exponentially activated
form with vanishing zero temperature value
due to the charge gap. For any doping the low-temperature
behavior is changed completely with finite value at zero temperature 
consistent with a partial filling of the lower Hubbard band.
For density $n=0.5$ we observe two different structures at low temperature
similar to the case of the specific heat.
The lower temperature ``spin'' peak resembles the structure in the 
susceptibility $\chi$, whereas the ``charge'' maximum at slightly 
higher temperature is caused by the single particle motion of the bare 
electrons.
The compressibility has a singular dependence on doping. The smaller the
doping the closer the curves are at high temperatures
and {\it the more divergent} at lower temperature, see
Fig.\ref{fig:hubConstU-K}. This, of course, is
exactly the behavior of a system exhibiting a Mott-Hubbard transition
at half-filling.

Our findings are qualitatively in accordance with the 
results\footnote{In \cite{KawUsuOki89,UsuKawOki90} 
notice the factor 4 for the interaction $U$.}
of \cite{KawUsuOki89,UsuKawOki90} for the dopings treated therein. 
In particular for specific heats, magnetic and charge
susceptibilities the results compare well for densities
$0.7\le n\le 1$ and temperatures $T\ge 0.1$, giving an independent support
to the truncation approximation adopted there.

The present approach has the advantage of  explicit  evaluations over much
wider temperature and density regions.
The $T$-linearity of the specific heat at very low temperatures,
as expected from CFT, is clearly observed in contrast to \cite{UsuKawOki90}.
Moreover,  we have novel observations of additional structures at lower 
temperatures and densities especially in the compressibility as mentioned
above. These structures can be also interpreted in terms of CFT and 
zero temperature excitations.  
Therefore, we conclude that the present approach is the first one
making possible the explicit evaluation of
the  crossover from the very low temperature (CFT) to the very high 
temperature region in an exact way.

In Fig.\ref{fig:hubSpinCharge} we show a separation 
of the specific heat into spin and charge
components. This is done in principle on the basis 
of eigenvalue equations like (\ref{nle-eval2-app}). As motivated by the 
study of the strong-coupling limit in section 7.1,
contributions by $\b$ and $\c$ functions
are interpreted as spin and charge contributions, respectively. However,
the procedure is not unique as we have various alternative formulations
resulting in different separations. In particular we like to note the
expression
\begin{eqnarray}
\ln \Lambda = -\beta(e_{0} -U/4 -\mu)  
&+& \int_{-1}^1 \left[c_0 \Delta \ln \C/\bC
- {\cal K} \ln(1+\c^+  +\bc^+)(1+\c^-  +\bc^-)\right]dx\nonumber\\
&+&  \int_{-\infty}^\infty  c_2 (x) \ln \B^-(x) dx  
+  \int_{-\infty}^\infty c_1(x) \ln \bB^+(x) dx,
\label{sep}
\end{eqnarray}
with
\begin{equation}
c_0(x)= \frac{1}{2\pi} \int_{-\infty}^\infty 
\frac{J_0(k)}{1+e^{U|k|/2}} e^{ikx}dk,\qquad
c_{1,2}(x)= \frac{1}{2\pi} \int_{-\infty}^\infty 
\frac{J_0(k)}{1+e^{\mp Uk/2}} e^{ikx}dk.
\end{equation}
Here $e_0$ is the groundstate energy at half-filling as given in
\cite{LiebWu68} and the additional
contributions by $\b$ and $\c$ functions represent correction terms
due to spin and charge excitations. In Fig.\ref{fig:hubSpinCharge} 
we show the results for the specific heat
\begin{equation}
c=T\left(\frac{\partial S}{\partial T}\right)_\mu+
T\left(\frac{\partial n}{\partial T}\right)_\mu
\left(\frac{\partial \mu}{\partial T}\right)_n,
\end{equation}
where we have applied the separation based on (\ref{sep}) to the 
temperature derivatives of $S$ and $n$.
Note the functional form of the spin part
which is rather independent of the doping. However, upon small doping
the charge contribution develops a low-temperature peak which disappears
again for larger dopings. 
We would like to comment on the issue of the ``mathematical
separation'' of spin and charge as described above (and similarly applied in
\cite{KawUsuOki89,UsuKawOki90}) that it may give rise to artificial results.
For instance, at higher temperatures the
``partial specific heats'' show negative values 
whereas the total specific heat is always positive. 
In section~\ref{sec:low-temperature-limit}
the spin-charge separation is treated at low temperatures
and arbitrary particle density via an involved interplay
of the various degrees of freedom rather than by a superficial interpretation
of formulas. 

\section{Analytical solutions of the integral equations}
\label{sec:analytical-solutions}

In the previous sections we have derived non-linear integral equations
for the largest eigenvalue being directly related to the free energy
of the Hubbard model at finite temperatures $T=1/\beta$. For arbitrary
temperatures and densities the integral equations can be solved only 
numerically. However, in some limiting cases analytical results can be 
derived and relations obtained 
which permit a comparison to known analytical results. This implies
the consistency of our approach.

\subsection{Strong-coupling limit}
\label{sec:strong-coupling-limit}

In the strong-coupling limit \mbox{$U\to\infty$} at half-filling
($\mu=0$) the Hubbard model is expected to reduce to the
Heisenberg chain. Indeed, in the strong-coupling limit 
we find directly the thermodynamics of the Heisenberg model. This can
be seen as follows:

Considering the limit $\gamma\to\infty$ (with $\gamma=U/4$)
we rescale the
argument of the auxiliary functions by \mbox{$x\mapsto 2\gamma x$}
as well as the ratio 
$\beta/(2\gamma)\mapsto\tilde{\beta}$ and
$(2\gamma)H\mapsto\tilde{H}$.
It turns out that all contributions of ${\frak c}^\pm$
and $\overline{{\frak c}^\pm}$ can be dropped in (\ref{nle})
because of their vanishing range of integration. 
Moreover ${\frak c}^\pm$
and $\overline{{\frak c}^\pm}$ tend to zero. 
The remaining equations read
\begin{equation}
  \log{\frak b}^{\pm}=
  \Psi_{\frak b}^\pm
  -K_{2,\pm\alpha-\alpha}\ast\log{\frak B}^{+}
  +K_{2,\pm\alpha+\alpha}\ast\log{\frak B}^{-}
  \quad{\rm with}\quad
  \Psi_{\frak b}^\pm=-\tilde\beta\tilde H +
2\pi\tilde{\beta}\,\overline{K}_{1,\pm\alpha}.
\end{equation}
According to equation (\ref{nle-eval2-app}) and after dropping the
groundstate energy shift $\beta U/4$ the eigenvalue is
\begin{equation}
  \log\Lambda= K_{1,-\alpha}\ast\log{\frak B}^{+}|_{x=0} 
              -K_{1,\alpha}\ast\log{\frak B}^{-}|_{x=0} 
+\tilde\beta\tilde H/2.
\end{equation}
Now we define 
\begin{equation}
  b:=1/{{\frak b}^+},\quad B:=1+b\quad{\rm and}\quad
  \overline{b}:={\frak b}^-,\quad \overline{B}:={\frak B}^{-},
\label{B-relations-in-large-U-limit}
\end{equation}
which are inserted into the integral equations. By means of the identity
\begin{equation*}
  \log{\frak B}^+=\log B-\log b,
\end{equation*}
we have
\begin{align*}
  -\log{b}-K_{2,0}\ast\log{b} &=
  +\Psi_{\frak b}^+
  -K_{2,0}\ast\log{B}
  +K_{2,2\alpha}\ast\log\overline{B},\\
  -\log\overline{b}-K_{2,0}\ast\log\overline{b} &=
  -\Psi_{\frak b}^-
  +K_{2,-2\alpha}\ast\log{B}
  -K_{2,0}\ast\log\overline{B}.
\end{align*}
Using the Fourier transform this provides 
\begin{align}
  \log{b} &=
 +\tilde\beta\tilde H/2 -2\pi\tilde{\beta}\,\Phi_{+\alpha}
  +R_{0}\ast\log{B}
  -R_{+2\alpha}\ast\log\overline{B},\label{nle-large-U1}\\
  \log\overline{b} &=
 -\tilde\beta\tilde H/2 -2\pi\tilde{\beta}\,\Phi_{-\alpha}
  +R_{0}\ast\log\overline{B}
  -R_{-2\alpha}\ast\log{B},\label{nle-large-U2}  \\
\intertext{with the eigenvalue}
  \log\Lambda &=
  2\tilde{\beta}\log{2}+
  \int_{-\infty}^{\infty}
    \left(\,\Phi_{\alpha}\log{B}-\Phi_{-\alpha}\log\overline{B}\,\right)\,
    {\rm d}x,\label{nle-eval-large-U}   \\
\intertext{and}
  \Phi_\alpha &=\frac{i}{2\sinh\pi(x+i\alpha)},\quad
  R_\alpha= \int_{-\infty}^{\infty}\frac{{\rm d}k}{2\pi}\,
             \frac{{\rm e}^{ikx-k\alpha}}{1+{\rm e}^{|k|}}.\nonumber
\end{align} 
These relations correspond to the non-linear integral equations of
the isotropic antiferromagnetic spin-$1/2$ Heisenberg chain
\cite{Klu92,Klu93}. In addition, this case is also related to the
thermodynamics of the $t-J$
model at half-filling \cite{JutKluSuz97}.

\subsection{Free-Fermion limit}
\label{sec:free-fermion-limit}

Let us consider the  opposite limit $U\to{}0$ (at $\mu=0$ and $H=0$),  
that is,  the case  of two
independent Free-Fermion  systems.  The non-linear integral equations
(\ref{nle2-app}) simplify to an algebraic set of equations 
\begin{align*}
  \log{\frak b}^{\pm}&=-\beta H
  -\log{\frak B}^{+}
  +\log{\frak B}^{-}
  -(\textstyle\frac{1}{2}\pm\textstyle\frac{1}{2})\,
      \Delta\log(\overline{{\frak c}}/\overline{{\frak C}}),\\
  \log{\frak c}^\pm&= +\beta(\mu+H/2)
  +\log\phi_\pm
  -\log\overline{{\frak B}}^-
  +(\textstyle\frac{1}{2}\pm\textstyle\frac{1}{2})\,
       \Delta\log\overline{{\frak C}},\\
  \log\overline{{\frak c}^\pm}&= -\beta(\mu+H/2)
  -\log\phi_\pm
  -\log{{\frak B}^+}
  +(-\textstyle\frac{1}{2}\pm\textstyle\frac{1}{2})\,
       \Delta\log{{\frak C}},  \\
\intertext{with}
  \log\Lambda&=
    -\int_{-1}^{1}{\cal K}_0\,
         \log\frac{(1+{\frak c}^++\overline{{\frak c}^+})
                   (1+{\frak c}^-+\overline{{\frak c}^-})
                   (1+{\frak c}^-)}
                  {\overline{{\frak c}^+}\,\overline{{\frak c}^-}\,
                   (1+{\frak c}^+)}\,{\rm d}x.
\end{align*}
The solution reads as follows
\begin{align*}
{\frak b}^{+}&=\frac{\left[1+\e^{\beta(\mu+H/2)}\phi\right]
\left[1+\e^{\beta(\mu-H/2)}/\phi\right]}
{\e^{\beta H}\left[
{\e^{\beta(\mu+H/2)}/\phi+\e^{2\beta\mu}+1+\e^{\beta(\mu-H/2)}/\phi}\right]},\\
{\frak b}^{-}&=\frac{\left[
{\e^{\beta(\mu+H/2)}/\phi+\e^{2\beta\mu}+1+\e^{\beta(\mu-H/2)}/\phi}\right]}
{\e^{\beta H}\left[1+\e^{\beta(\mu+H/2)}/\phi\right]
\left[1+\e^{\beta(\mu-H/2)}\phi\right]},\\
\c^+&=\frac{\e^{\beta(\mu+H/2)}}{\phi}
\frac{1+\e^{\beta(\mu-H/2)}\phi}
{1+\e^{\beta(\mu-H/2)}/\phi}\frac{\b^+}{1+\b^+},\qquad
\c^-=\frac{\e^{\beta(\mu+H/2)}}{\phi}\frac{\b^-}{1+\b^-},\\
\bc^+&=\frac{1}{\e^{\beta(\mu+H/2)}\phi(1+\b^+)},\qquad
\bc^-=\frac{\phi}{\e^{\beta(\mu+H/2)}}
\frac{1+\e^{\beta(\mu-H/2)}/\phi}
{1+\e^{\beta(\mu-H/2)}\phi}\frac{1}{1+\b^-},
\end{align*}
Lastly, we substitute $x=\sin{k}$ in the integration for the
eigenvalue which leads to
\begin{align*}
  \log\Lambda=&
     +\frac{1}{2\pi}\int_{-\pi}^{\pi}
\ln\left[1+\exp\left(\beta(\mu+H/2+2\cos{k})\right)\right]\,{\rm d}k\\
     &+\frac{1}{2\pi}\int_{-\pi}^{\pi}
\ln\left[1+\exp\left(\beta(\mu-H/2+2\cos{k})\right)\right]\,{\rm d}k.
\end{align*}
This is the desired result.

\subsection{Low-temperature asymptotics}
\label{sec:low-temperature-limit}

The  low-temperature regime is the most interesting limit as the system
shows Tomonaga-Luttinger liquid behavior. We will derive analytic
expressions for the thermodynamics within our first principles calculations
and confirm the field theoretical predictions. In particular we will show
how the 
non-linear integral  equations correspond to  the known dressed energy
formalism of the Hubbard model. This represents a further and in fact
the most interesting consistency check. 

For $T=1/\beta\ll{}0$ we  can simplify the non-linear
integral equations as follows.   We adopt fields $H>0$, $\mu<0$,   such
that ${\frak b}^-\to{}0$, ${\frak c}^\pm\to{}0$ and
$1/\overline{{\frak c}}^-\to{}0$ at
$\beta\gg{}1$ which can be verified numerically. Moreover, one finds
\begin{equation*}
  |{\frak b}^+|,\, |1/\overline{{\frak c}}^+|\gg{}1
                 \quad{\rm for}\quad |x|<\lambda_0,\, {\sigma}_0    
  \quad{\rm and}\quad
  |{\frak b}^+|,\, |1/\overline{{\frak c}}^+|
                 \ll{}1\quad{\rm for}\quad |x|>\lambda_0,\, {\sigma}_0,
\end{equation*}
for certain crossover values $\lambda_0,\, {\sigma}_0$.
The slopes for the crossover are steep, so that 
the following approximations to the integral equations (\ref{nle2-app})
are valid:
\begin{equation}
\begin{split}
 \log{\frak b}^+ &= \phi_b
        -\int_{-\lambda_0}^{+\lambda_0} 
             K_{2}(\lambda-\lambda')\, \log {\frak b}^+ (\lambda')\,
             {\rm d}\lambda'
        +\int_{-k_0}^{+k_0} 
             \overline{K}_{1,\alpha}(\lambda-\sin{k'})\,
           \cos{k'}\, \log {\frak c}^{\vee}(k') \,
             {\rm d}k'\\
 \log {\frak c}^{\vee} &= \phi_c
        +\int_{-\lambda_0}^{+\lambda_0} 
             K_{1,-\alpha}(\sin{k}-\lambda')\,\log {\frak b}^+(\lambda')\,
             {\rm d}\lambda'. 
\label{linearbc}
\end{split}
\end{equation}
Here we use ${\frak c}^{\vee}(k)= 1/\overline{{\frak c}}^+(\sin k)$
and ${\sigma}_0= \sin k_0$. 
The driving terms read 
\begin{align}
\phi_b(\lambda) &= -\beta \varepsilon_s^0(\lambda) 
       -\frac{\pi^2 (K_{2}(\lambda-\lambda_0)+K_{2}(\lambda+\lambda_0))}
             {6 (\log 1/{\frak b}^+)'(\lambda_0)}\nonumber\\
&\phantom{= -\beta \varepsilon_s^0(\lambda)  }
       +\frac{\pi^2 \cos k_0 (K_{1,\alpha}(\lambda-\sin k_0)
+K_{1,\alpha}(\lambda+\sin k_0))}
            {6 (\log 1/{\frak c}^{\vee})'(k_0)},\nonumber\\
\phi_c(\lambda) &= -\beta \varepsilon_c^0(\lambda) + 
     \frac{\pi^2 (K_{1,-\alpha}(\sin k-\lambda_0)+
                  K_{1,-\alpha}(\sin k+\lambda_0))}
             {6 (\log 1/{\frak b}^+)'(\lambda_0)},
\label{expans}
\end{align}
where $\varepsilon_s^0=H$, 
$\varepsilon_c^0= -\mu -U/2-H/2- 2 \cos{k}$.
Retaining only the leading terms in the integral equations
and choosing the imaginary part of the integration contour
as $\alpha=\gamma$, 
we find the following connections between
auxiliary functions and the dressed energy functions:
\begin{equation}
 \log{\frak b}^+ =- \beta\,\varepsilon_s+ O(1/\beta) \quad{\rm and}\quad
 \log {\frak c}^{\vee}=- \beta\,\varepsilon_c+ O(1/\beta).
\label{BC-relations-in-low-T-limit}
\end{equation}
For a comparison with \cite{FraKor90,FraKor91} note the different normalization
of the chemical potential.
The free energy also admits the same approximation scheme yielding
up to $O(T^2)$ terms in the low-temperature expansion.
To present this, we introduce ``root density functions'' $\rho$ by the
definition
\begin{equation}
\begin{split}
 \rho_s (\lambda) &= 
        -\int_{-\lambda_0}^{+\lambda_0} 
             K_{2}(\lambda-\lambda')\,\rho_s (\lambda')\,
             {\rm d}\lambda'    
       +\int_{-k_0}^{+k_0} 
             \overline{K}_{1,\alpha}(\lambda-\sin{k'})\,
            \rho_c (k') \, {\rm d}k'\\
 \rho_c(k) &= \frac{1}{2\pi} 
        + \cos k 
          \int_{-\lambda_0}^{+\lambda_0} 
             K_{1,-\alpha}(\sin{k}-\lambda')\, \rho_s(\lambda')\,
             {\rm d}\lambda'. \\
\label{rhosc}
\end{split}
\end{equation}
Note that the kernel matrices for the integral equations
(\ref{linearbc}, \ref{rhosc}) are mutually transpose.
The following equality is an immediate consequence:

\begin{equation}
\frac{1}{2\pi} \int_{-k_0}^{+k_0}  \log {\frak c}^{\vee} (k) {\rm d}k
      =
               \int_{-k_0}^{+k_0}\rho_c(k)\,\phi_c(k)\,{\rm d}k
                +\int_{-\lambda_0}^{+\lambda_0} 
                 \rho_s(\lambda)\,\phi_b(\lambda)\,{\rm d}\lambda.
\end{equation}
With these relations the eigenvalue $(\log\Lambda)$  reads
\begin{equation}
  \log\Lambda =-\beta\gamma+
               \int_{-k_0}^{+k_0}\rho_c(k)\,\phi_c(k)\,{\rm d}k
                +\int_{-\lambda_0}^{+\lambda_0} 
                 \rho_s(\lambda)\,\phi_b(\lambda)\,{\rm d}\lambda
              +\frac{\pi}{6 \beta \varepsilon_c'(k_0)},
\label{semi-fin}
\end{equation}
where we have replaced $(\log 1/{\frak c}^{\vee})'(k_0)$
in the denominator by $\beta \varepsilon_c'(k_0)$.
Substituting (\ref{expans}) into (\ref{semi-fin}) we arrive at
the final expression,
\begin{equation}
f=\varepsilon_0 -\frac{\pi}{6\beta^2}\,
\left(\frac{1}{v_c}+\frac{1}{v_s}\right).
\label{lowTsep}
\end{equation}
The definitions of sound velocity and the ground state 
energy coincide with standard results,
$v_{s,c}=\varepsilon'_{s,c}/2\pi\rho_{s,c}|_{\lambda_0, k_0}$, and
\begin{equation*}
\varepsilon_0 =\int_{-k_0}^{+k_0}\rho_c(k)\,\epsilon^0_c(k)\,{\rm d}k
                +\int_{-\lambda_0}^{+\lambda_0}
                 \rho_s(\lambda)\,\epsilon^0_s(\lambda)\,{\rm d}\lambda.
\end{equation*}
Here the trivial shift in the energy $U/4$ is omitted.
We thereby conclude that our formalism completely recovers
the correct contribution from spinon and holon
excitations in the low temperature behavior. 
This is a manifestation of spin-charge separation due to which each
elementary excitation contributes independently to (\ref{lowTsep})
where the velocities $v_c$ and $v_s$ typically take different values.

\subsection{High-temperature limit }

Finally, we consider the high-temperature limit
$T\to\infty$ with $H,U$ as well as $\beta\mu$ fixed. 
The auxiliary functions in (\ref{nle}) become constant
\begin{align*}
 \log{\frak b}^\pm& = 0,\quad
 \log{\frak c}^\pm  =  \beta\mu-\log{2},\quad
 \log\overline{{\frak c}^\pm} = -\beta\mu-\log{2},\\
 \hookrightarrow\quad\log\Lambda&=
   \log(1+{\frak c}+\overline{{\frak c}})/\overline{{\frak c}}.
\end{align*}
Thus, the free energy reads 
\begin{equation}
  f=-2\,T\log(1+{\rm e}^{\mu/T})\quad{\rm with}\quad
  \mu/T=\log\frac{n}{2-n},
\end{equation}
where $n$ is the particle density. We obtain the entropy
\begin{equation}
  S=2\log\frac{2}{2-n}-n\log\frac{n}{2-n},
\end{equation}
as expected by counting the
degrees of freedom per lattice site. Especially, 
at half-filling $n=1$ this equals to $S=\log(4)$.

\section{Summary and Discussion}

In this paper, the novel formulation of thermodynamics for 1D quantum
systems has been successfully applied to the Hubbard model.
Several quantities of physical interests have been
evaluated with high numerical precision and various limiting cases have
been studied analytically.

As already noted above, we may consider as one of the most practical 
advantages of the present formulation
the fact that one only has to deal with a finite number of 
unknown functions and nonlinear integral equations among them.
This does not only imply convenience, rather it opens a more 
fundamental understanding related to the particle picture of
1D quantum systems.
For the Heisenberg model, the complex conjugate auxiliary functions
play a role which seems to correspond to the elementary 
spinon excitations \cite{CloPea62,JohKriMcCoy73,FadTak81},
while for the integrable $t-J$ model one further function related
to the holon is needed.
In this paper, we have shown that three independent functions $\b,
\c, \bc$ describe the complete thermodynamics, physically 
corresponding to spinons, and holons in upper and lower
Hubbard bands. 
In the $T\rightarrow 0$ limit, these functions are shown to reduce
to energy density functions (``dressed energy functions'')
for such elementary excitations.

This interpretation which is natural at low temperatures
poses however a problem at finite temperatures.
The auxiliary functions, related to energies of
excitations at $T=0$, are no longer real for
arbitrary temperature.  
Thus they lose the direct connection to physical excitations in
the sense of energy levels. On the other hand, imaginary parts of energies
indicate a finite life-time of excitations, or in this case a decay of
the elementary particles of the system.
We leave the investigation of these questions as an interesting 
future problem.

Obviously, our formulation can be extended to the evaluation of 
the asymptotics of correlation functions, such as spin-spin 
correlation lengths etc.
These investigations on highly correlated electron systems,
including the Hubbard, the integrable $t-J$, supersymmetric $U$ models
will be reported together in the near future.

\section*{Acknowledgements}

The authors  acknowledge  financial   support  by the   {\it  Deutsche
Forschungsgemeinschaft} under grant  No.   Kl~645/3-1 and support by
the research program of the 
Sonderforschungsbereich 341, K\"oln-Aachen-J\"ulich.
\newpage
\begin{appendix}

\section{Derivation of integral expressions for the eigenvalue}
\label{sec:nlie-derivation}
Here we turn to the derivation of expressions for the largest eigenvalue
of the QTM in terms of the above auxiliary functions. We calculate
$\sum_j\ln z_j$ by a Cauchy integral of the function
$f(s)=\ln z(s)\left[\ln\left(1+{l_4}/{l_3}(s)\right)\right]'$ where
the zeros of $(1+{l_4}/{l_3}(s))$ in the neighborhood of the real
axis are precisely the $s_j$. Furthermore
we use a contour \L{0} surrounding the $s_j$ in anticlockwise manner.
The $s_j$ are not located on the branch cut of $\ln z(s)$ from $-1$ to $1$,
hence \L{0} consists of two disconnected parts. (For not too low temperatures
and vanishing external fields these parts are loops around $]-\infty,-1]$
and $[1,\infty[$, respectively. In the general case they are appropriately
deformed.) However the $z_j$ corresponding to a particular $s_j$ might have 
to be calculated by the use of the second branch of $\ln z(s)$.
We therefore obtain
\begin{equation}
  \begin{split}
2\pi i\sum_j\ln z(s_j)&=
\int_{{\cal L}_0}f(s)\big|_{\rm 1. branch}ds
+\int_{{\cal L}_0}f(s)\big|_{\rm 2. branch}ds,\\
  \end{split}
\label{eig0}
\end{equation}
where the first and second term on the right hand side will be abbreviated
by $\Sigma_1$ and $\Sigma_2$, respectively. We next manipulate $\Sigma_1$
\begin{equation}
  \begin{split}
\Sigma_1&=
\int_{{\cal L}_0\equiv-({\cal L}_1+{\cal L}_2+{\cal L}_3)}
\ln z(s)\left[\ln\left(1+\frac{l_4}{l_3}(s)\right)\right]'ds,\\
&=-\int_{{\cal L}_1+{\cal L}_3}\ln z(s)
\left[\ln\left(1+\frac{l_4}{l_3}(s)\right)\right]'ds
+\int_{{\cal L}_2}\ln z(s)\left[\ln\bT(s)\right]'ds,
  \end{split}
\label{eig1}
\end{equation}
where in the first line the integration contour \L{0} can be replaced due
to Cauchy's theorem by three contours (taken in anticlockwise manner): \L{1} 
from $i\infty$ to $-1$, surrounding $[-1,1]$, from $-1$ back to
$i\infty$; \L{2} surrounding the axis $\Im(s)=-\gamma$ (where the simple 
poles of $[1+{l_4}/{l_3}(s)]$ are located, which are identical to the 
simple zeros of $\bT(s)$); \L{3} around $s_0$ (which is a pole of order $N/2$
of $[1+{l_4}/{l_3}(s)]$), see (\ref{bare}). 
Next we replace \L{2} by contours \L{1},
\L{{}}$-2i\gamma$ (where \L{{}} surrounds the real axis), and \L{3}$-2i\gamma$
\begin{equation}
  \begin{split}
\int_{{\cal L}_2}\ln z(s)\left[\ln\bT(s)\right]'ds=&
-\int_{{\cal L}_1}\ln z(s)\left[\ln\bT(s)\right]'ds\\
&-\int_{{\cal L}_3+{\cal L}}\ln z(s-2i\gamma)
\left[\ln\bT(s-2i\gamma)\right]'ds.
  \end{split}
\label{eig2}
\end{equation}
In the last term of (\ref{eig2}) the function $\ln\bT(s-2i\gamma)$ can be
replaced by $-\ln\B(s)$ as the difference of these functions amounts to
an analytic contribution vanishing in the contour integration. Of course,
the \L{3} integration in (\ref{eig1}) and (\ref{eig2}) can be done explicitly 
yielding
\begin{equation}
  \begin{split}
\Sigma_1&=\pi i N \ln [z(s_0)z(s_0-2i\gamma)]\\
&-\int_{{\cal L}_1}\ln z(s)
\left[\ln\left(\bT(s)\left(1+\frac{l_4}{l_3}(s)\right)\right)\right]'ds
+\int_{{\cal L}}\ln z(s-2i\gamma)\left[\ln\B(s)\right]'ds.
  \end{split}
\label{eig3}
\end{equation}
Next we perform integration by parts on the right hand side where the first
integral also contributes a ``surface term''
\begin{equation}
  \begin{split}
\Sigma_1&=
-2\pi i\ln\left[\left(1+\e^{-\beta(\mu+H/2)}\right)
\left(1+\e^{\beta(\mu-H/2)}\right)\right]+
\pi i N \ln [z(s_0)z(s_0-2i\gamma)]\\
&+\int_{{\cal L}_1}[\ln z(s)]'
\ln\left[\bT(s)\left(1+\frac{l_4}{l_3}(s)\right)\right]ds
-\int_{{\cal L}}[\ln z(s-2i\gamma)]'\ln\B(s)ds,
  \end{split}
\label{eig4}
\end{equation}
where now the integration along \L{1} can be 
restricted to a loop \L{4} along the cut from -1 to 1. In the limit
$N\to\infty$ we can replace $\pi i N \ln [z(s_0)z(s_0-2i\gamma)]$
by $2\pi i N \ln z_-(w)+4\pi i\beta\gamma$ such that in combination
with (\ref{eig_prac})
\begin{equation}
  \begin{split}
2\pi i\ln\Lambda=2\pi i\beta(\mu+H/2+U/4)+\Sigma_2&+
\int_{{\cal L}_4}[\ln z(s)]'
\ln\left[\bT(s)\left(1+\frac{l_4}{l_3}(s)\right)\right]ds\\
&-\int_{\cal L}[\ln z(s-2i\gamma)]'\ln\B(s)ds.
  \end{split}
\label{eig5}
\end{equation}
A very similar line of reasoning yields
\begin{equation}
  \begin{split}
2\pi i\ln\Lambda=2\pi i\beta(\mu-H/2+U/4)+\Sigma_2'&+
\int_{{\cal L}_4}[\ln z(s)]'
\ln\left[\T(s)\left(1+\frac{l_3}{l_4}(s)\right)\right]ds\\
&-\int_{\cal L}[\ln z(s+2i\gamma)]'\ln\bB(s)ds,
  \end{split}
\label{eig5b}
\end{equation}
where $\Sigma_2'$ is defined similarly to $\Sigma_2$ after interchanging
$l_3$ and $l_4$.
Combining (\ref{eig5},\ref{eig5b}) we find the symmetrised version
\begin{equation}
  \begin{split}
4\pi i\ln\Lambda=&4\pi i\beta(\mu+U/4)+\Sigma_2+\Sigma_2'\\
&+\int_{{\cal L}_4}[\ln z(s)]'
\ln\left[\T(s)\bT(s)\left(1+\frac{l_4}{l_3}(s)\right)
\left(1+\frac{l_3}{l_4}(s)\right)\right]ds\\
&-\int_{\cal L}[\ln z(s-2i\gamma)]'\ln\B(s)ds
-\int_{\cal L}[\ln z(s+2i\gamma)]'\ln\bB(s)ds.
  \end{split}
\label{eig6}
\end{equation}
The present formula is still inconvenient as the first terms on the
right hand side contain ``non-standard'' functions. However, we can
substitute the terms 
\begin{equation}
\left[\left(1+\frac{l_4}{l_3}(s)\right)
\left(1+\frac{l_3}{l_4}(s)\right)\right]\Bigg|_{\rm 1. branch}
\longrightarrow\left[\left(1+\frac{l_4}{l_3}(s)\right)
\left(1+\frac{l_3}{l_4}(s)\right)\right]\Bigg|_{\rm 2. branch}
\label{eig7}
\end{equation}
without change of the integral as $[\ln z(s)]'|_{\rm 1. branch}
=-[\ln z(s)]'|_{\rm 2. branch}$. For the same reason we find
\begin{equation}
  \begin{split}
\Sigma_2+\Sigma_2'&=
-\int_{{\cal L}_0}\left([\ln z(s)]'
\ln\left[\left(1+\frac{l_4}{l_3}(s)\right)
\left(1+\frac{l_3}{l_4}(s)\right)\right]\right)\Bigg|_{\rm 2. branch}ds\\
&=
\int_{{\cal L}_0}[\ln z(s)]'
\ln\left[\left(1+\frac{l_4}{l_3}(s)\right)
\left(1+\frac{l_3}{l_4}(s)\right)\right]\Bigg|_{\rm 2. branch}ds,
  \end{split}
\label{eig1b}
\end{equation}
where functions have to be evaluated on the 1. branch unless indicated
differently. Inserting (\ref{eig7},\ref{eig1b}) into (\ref{eig6}),
combining the contours \L{0} and \L{4} into \L{{}}, and simply extending 
\L{4} to \L{{}} for the integrals involving $\T$ and $\bT$ (due to 
analyticity) we arrive at
\begin{equation}
  \begin{split}
4\pi i\ln\Lambda=&4\pi i\beta(\mu+U/4)\\
&+
\int_{{\cal L}}[\ln z(s)]'
\ln\left[\T(s)\bT(s)\left(\left(1+\frac{l_4}{l_3}(s)\right)
\left(1+\frac{l_3}{l_4}(s)\right)\right)\Bigg|_{\rm 2. branch}\right]ds\\
&-\int_{\cal L}[\ln z(s-2i\gamma)]'\ln\B(s)ds
-\int_{\cal L}[\ln z(s+2i\gamma)]'\ln\bB(s)ds.
  \end{split}
\label{eig6b}
\end{equation}
Next we find the identity
\begin{equation}
\left({l_4}/{l_3}\right)\big|_{\rm 2. branch}=
\left(\b{\T}/{\bT}\right)\big|_{\rm 1. branch}
\label{eig8}
\end{equation}
Hence the integrand of the first integral in (\ref{eig6b}) is 
\begin{equation}
  \begin{split}
\T\bT \left[\left(1+\frac{\b{\T}}{\bT}\right)
\left(1+\frac{\bT}{\b{\T}}\right)\right]=
\frac{(\bT+\b\T)^2}{b}=(1+\c+\bc)^2\B\bB,
  \end{split}
\label{eig9}
\end{equation}
with all functions on the first branch.
We are now in the position to formulate the first main expression for
the eigenvalue
\begin{equation}
  \begin{split}
4\pi i\ln\Lambda&=4\pi i\beta(\mu+U/4)+2\int_{\cal L}[\ln z(s)]'
\ln\left(1+\c+\bc\right)ds\\
&-\int_{\cal L}\left[\ln\frac{z(s-2i\gamma)}{z(s)}\right]'\ln\B(s)ds
-\int_{\cal L}\left[\ln\frac{z(s+2i\gamma)}{z(s)}\right]'\ln\bB(s)ds.
  \end{split}
\label{eig10}
\end{equation}
The last equation can be reduced further by substituting the
$\bB$ integral by a $\B$ integral. For this purpose we employ (\ref{auxFunct3})
\begin{equation}
  \begin{split}
\int_{\cal L}&\left[\ln\frac{z(s+2i\gamma)}{z(s)}\right]'\ln\bB(s)ds\\
&=
\int_{\cal L}\left[\ln\frac{z(s+2i\gamma)}{z(s)}\right]'
\ln(l_1+l_2+l_3+l_4+\bl{1}+\bl{2}+\bl{3}+\bl{4})ds\\
&-
\int_{\cal L}\left[\ln\frac{z(s+2i\gamma)}{z(s)}\right]'
\ln(\bl{1}+\bl{2}+\bl{3}+\bl{4})ds.
  \end{split}
\label{eig11}
\end{equation}
In the first term on the right hand side the contribution due to 
$z(s+2i\gamma)$ vanishes upon taking the contour integral and can
be replaced by (the equally vanishing) $z(s-2i\gamma)$. In the second term
the contour is replaced by $-($\L{{}}$-2i\gamma)$. Using
$(\bl{1}+\bl{2}+\bl{3}+\bl{4})(s-2i\gamma)=
\exp(-\beta H)(l_1+l_2+l_3+l_4)(s)/[{\phi^+(s)\phi^-(s)}]$ we find
\begin{equation}
  \begin{split}
\int_{{\cal L}\equiv-({\cal L}-2i\gamma)}\left[\ln\frac{z(s+2i\gamma)}{z(s)}\right]'
&\ln(\bl{1}+\bl{2}+\bl{3}+\bl{4})ds\\
&=2\pi i\beta H+\int_{\cal L}\left[\ln\frac{z(s-2i\gamma)}{z(s)}\right]'
\ln(l_1+l_2+l_3+l_4)ds,
  \end{split}
\label{eig12}
\end{equation}
where terms involving $\phi^+\phi^-$ have been dropped as they vanish in the
limit $N\to\infty$.
Inserting this into (\ref{eig11}) we obtain
\begin{equation}
\int_{\cal L}\left[\ln\frac{z(s+2i\gamma)}{z(s)}\right]'\ln\bB(s)ds
=-2\pi i\beta H+
\int_{\cal L}\left[\ln\frac{z(s-2i\gamma)}{z(s)}\right]'\ln\B(s)ds,
\end{equation}
and with (\ref{eig10}) we find
\begin{equation}
  \begin{split}
2\pi i\ln\Lambda=2\pi i\beta(\mu+H/2+U/4)&+\int_{\cal L}[\ln z(s)]'
\ln\left(1+\c+\bc\right)ds\\
&-\int_{\cal L}\left[\ln\frac{z(s-2i\gamma)}{z(s)}\right]'\ln\B(s)ds.
  \end{split}
\label{eig13}
\end{equation}

Finally, we want to show how to express the eigenvalue entirely in terms
of $\c$ functions, i.e. without contributions by $\b$. To this end we 
note the identity
\begin{equation}
  \begin{split}
\int_{{\cal L}\equiv-({\cal L}-2i\gamma)}\left[\ln{z(s)}\right]'\ln\bt(s)ds=&
-2\pi i\beta (\mu+H/2+U/2)\\
&-\int_{{\cal L}}\left[\ln{z(s-2i\gamma)}\right]'\ln\bt(s-2i\gamma)ds,
  \end{split}
\label{eig14}
\end{equation}
where we have dropped terms that do not contribute in the limit $N\to\infty$.
Next we replace the $\bt(s)$ and $\bt(s-2i\gamma)$ functions by
\begin{equation}
\bt=\bc\B, \qquad [\bt(s-2i\gamma)]^{-1}=\t(s)=\frac{l_1+l_2}
{l_3+l_4+\bl{1}+\bl{2}+\bl{3}+\bl{4}} \B\C.
\label{eig15}
\end{equation}
Since the function $({l_1+l_2})/({l_3+l_4+\bl{1}+\bl{2}+\bl{3}+\bl{4}})$
is analytic in the neighborhood of the real axis its contribution 
to (\ref{eig14}) vanishes. Therefore (\ref{eig14}) results into
\begin{equation}
  \begin{split}
\int_{{\cal L}}\left[\ln{z(s)}\right]'(\ln\bc(s)+\ln\B(s))ds=&
-2\pi i\beta (\mu+H/2+U/2)\\&+
\int_{{\cal L}}\left[\ln{z(s-2i\gamma)}\right]'(\ln\B(s)+\ln\C(s))ds.
  \end{split}
\label{eig16}
\end{equation}
Inserting this into (\ref{eig13}) we are left with
\begin{equation}
2\pi i\ln\Lambda=-2\pi i\beta\frac{U}{4}+\int_{\cal L}[\ln z(s)]'
\ln\frac{1+\c+\bc}{\bc}ds
+\int_{\cal L}\left[\ln{z(s-2i\gamma)}\right]'\ln\C(s)ds.
\label{eig17}
\end{equation}

\end{appendix}
\newpage

\newpage
\begin{figure}[t]
  \begin{center}
    \includegraphics[width=0.31\textwidth]{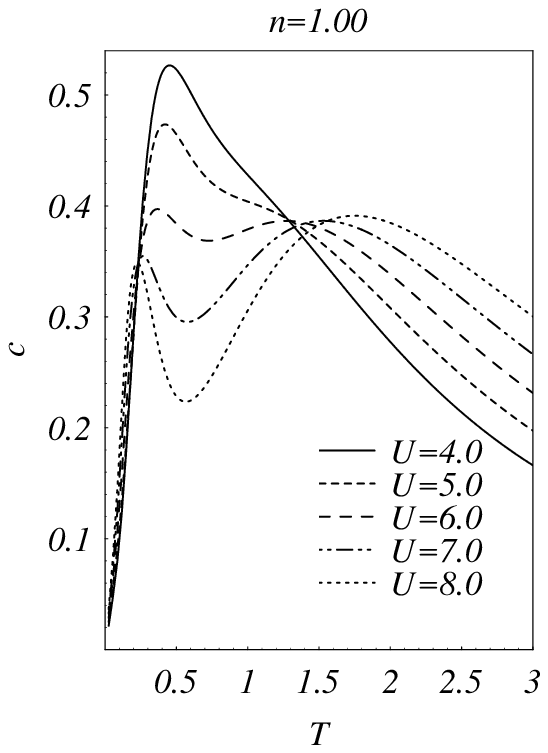}
    \includegraphics[width=0.31\textwidth]{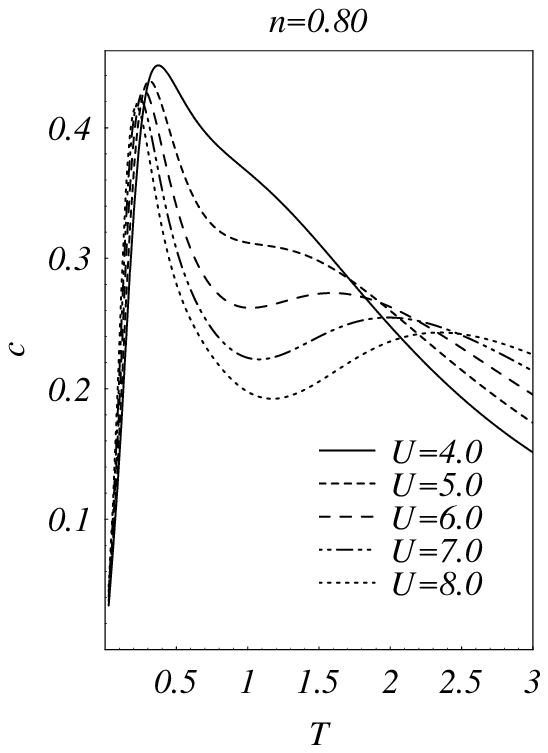}
    \includegraphics[width=0.31\textwidth]{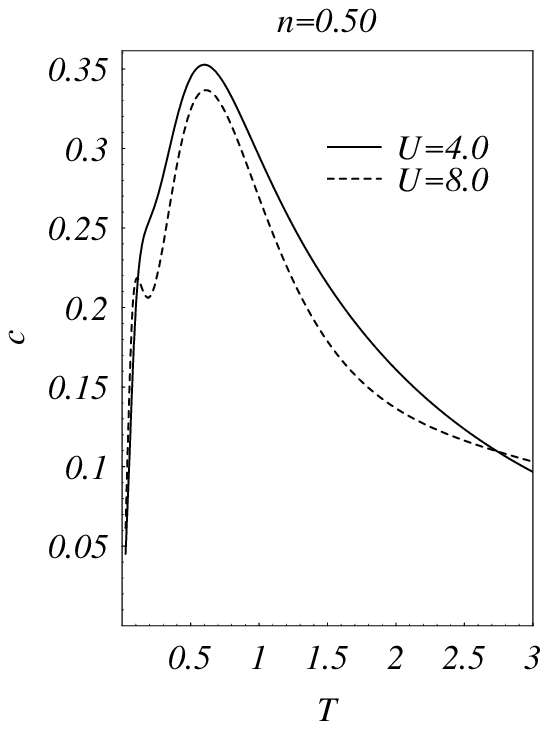}
  \end{center}
  \caption{Specific heat versus $T$ for particle densities
    $n=1$, $n=0.8$ and $n=0.5$.}
  \label{fig:hubU-C}
\end{figure}
\begin{figure}[b]
  \begin{center}
    \leavevmode  
    \includegraphics[width=0.31\textwidth]{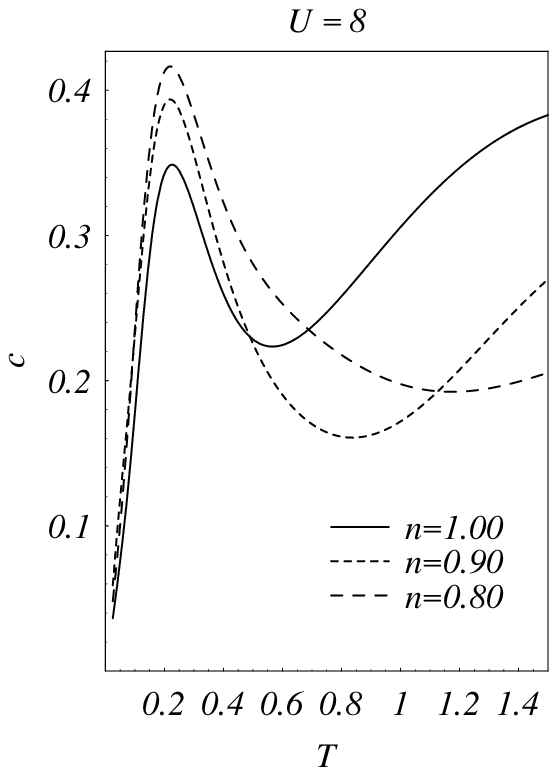}
    \includegraphics[width=0.31\textwidth]{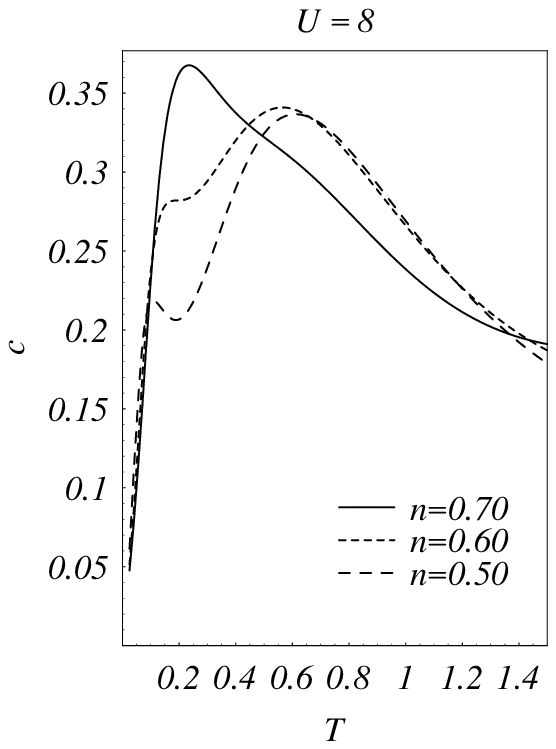}
    \includegraphics[width=0.31\textwidth]{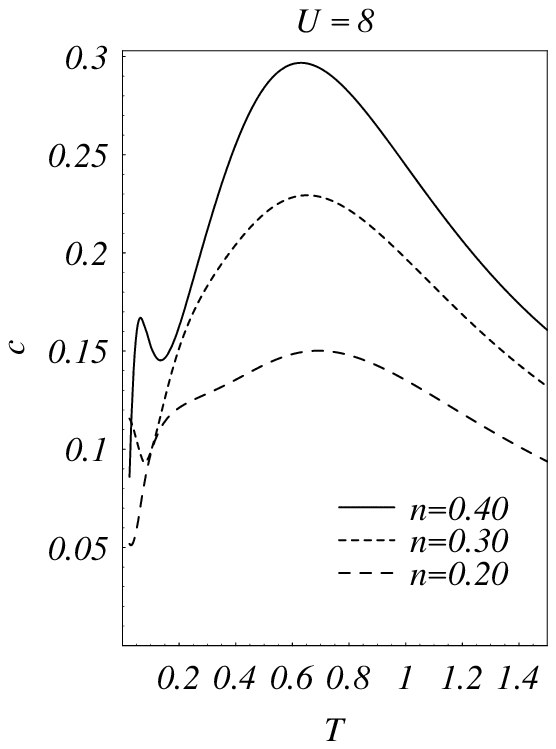}
  \end{center}
  \caption{Specific heat versus T for fixed $U=8$.}
  \label{fig:hubConstU-C}
\end{figure}
\begin{figure}[htb]
  \begin{center}
    \leavevmode  
    \includegraphics[width=0.31\textwidth]{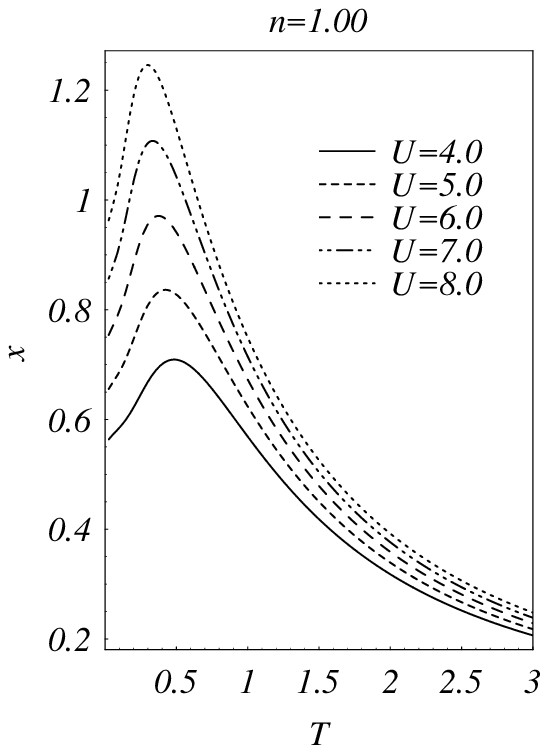}
    \includegraphics[width=0.31\textwidth]{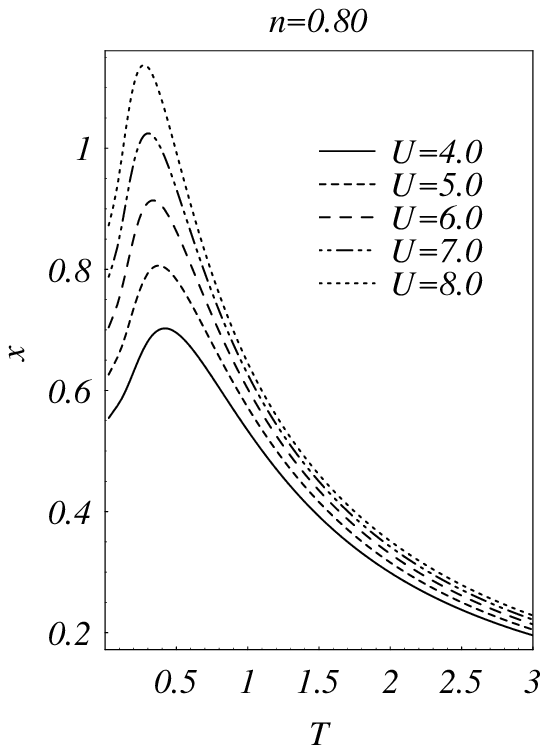}
    \includegraphics[width=0.31\textwidth]{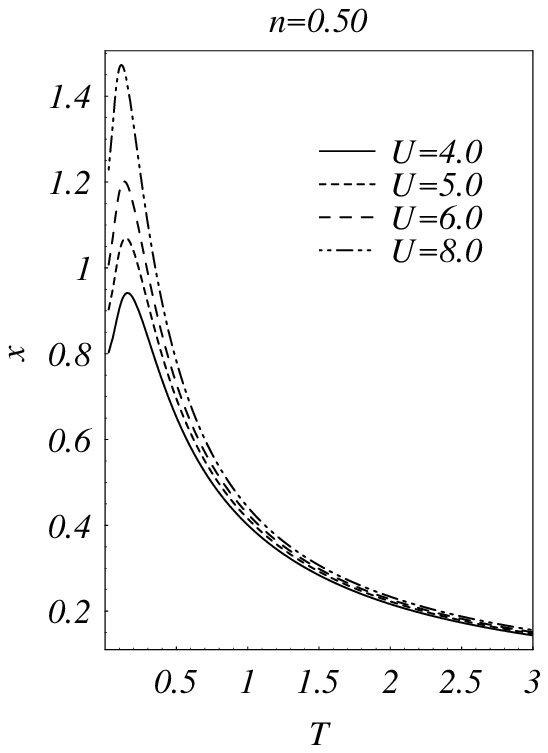}
  \end{center}
  \caption{Magnetic susceptibility versus $T$ for 
           $n=1$, $n=0.8$ and $n=0.5$.}
  \label{fig:hubU-X}
\end{figure}
\begin{figure}[htb]
  \begin{center}
    \leavevmode  
    \includegraphics[width=0.31\textwidth]{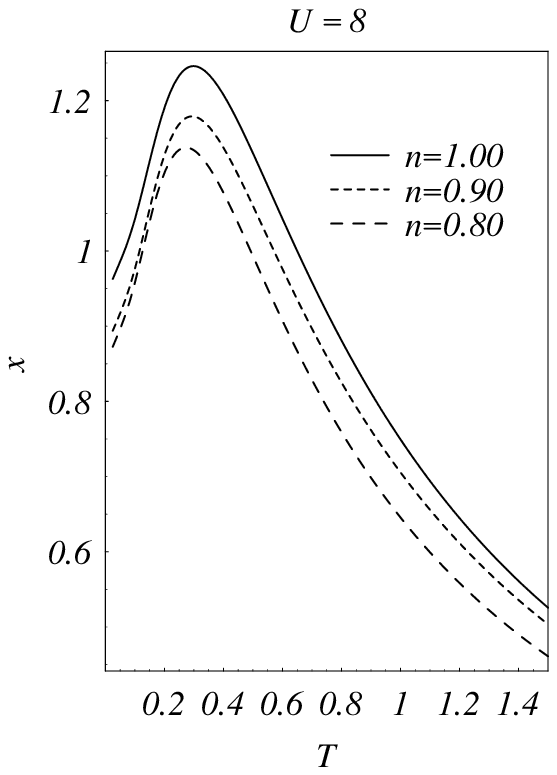}
    \includegraphics[width=0.31\textwidth]{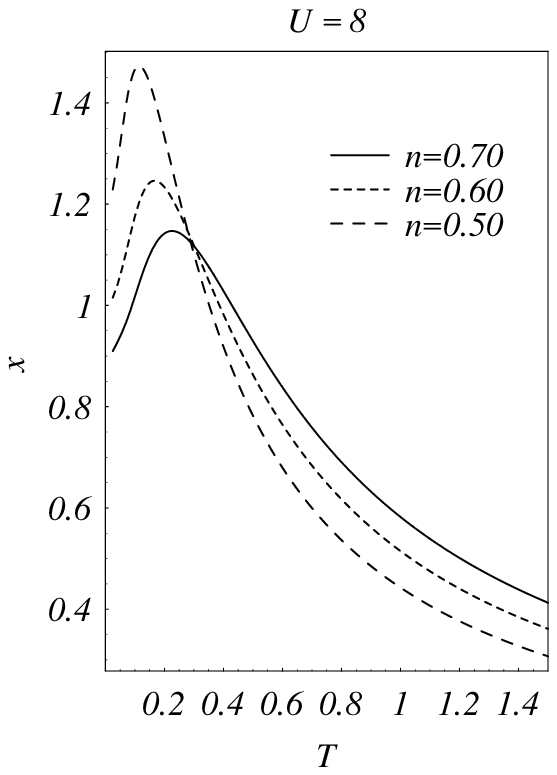}
    \includegraphics[width=0.31\textwidth]{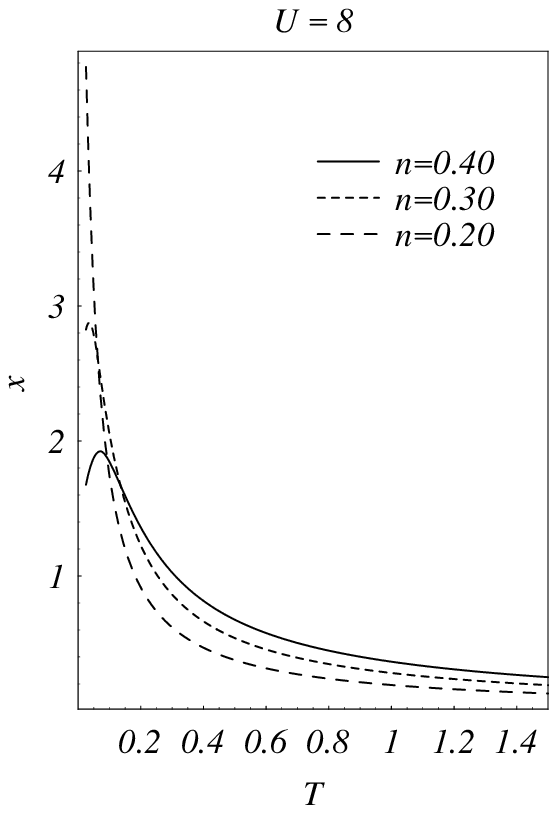}
  \end{center}
  \caption{Magnetic susceptibility versus T for fixed $U=8$.}
  \label{fig:hubConstU-X}
\end{figure}
\begin{figure}[h]
  \begin{center}
    \leavevmode  
    \includegraphics[width=0.31\textwidth]{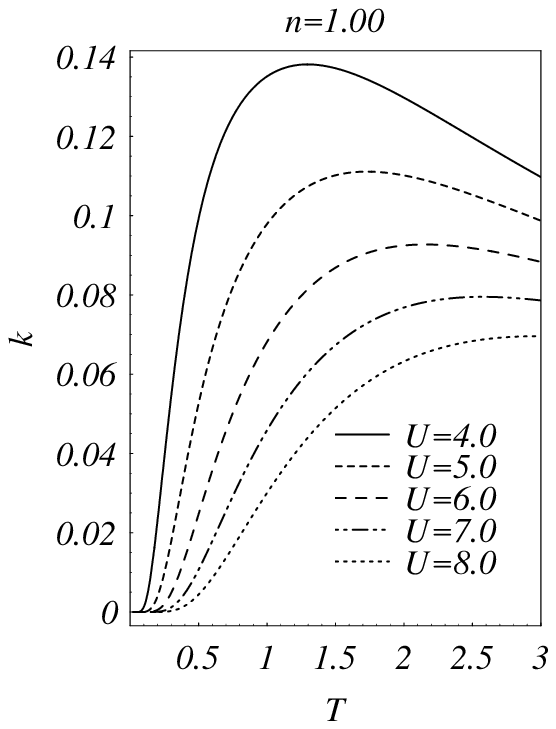}
    \includegraphics[width=0.31\textwidth]{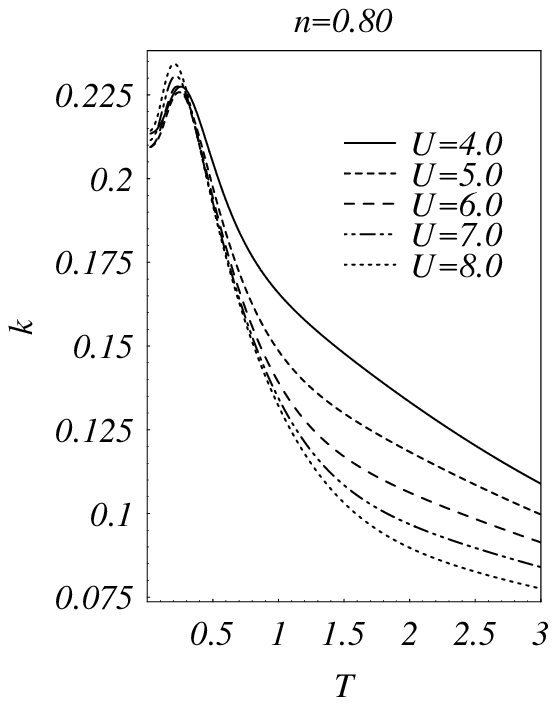}
    \includegraphics[width=0.31\textwidth]{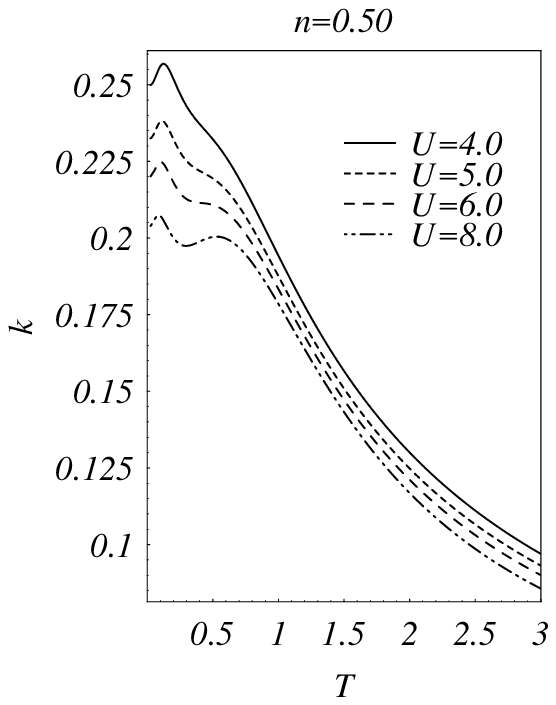}
  \end{center}
  \caption{Compressibility versus $T$ for particle densities
    $n=1$, $n=0.8$ and $n=0.5$.}
  \label{fig:hubU-K}
\end{figure}
\begin{figure}[htb]
  \begin{center}
    \leavevmode  
    \includegraphics[width=0.31\textwidth]{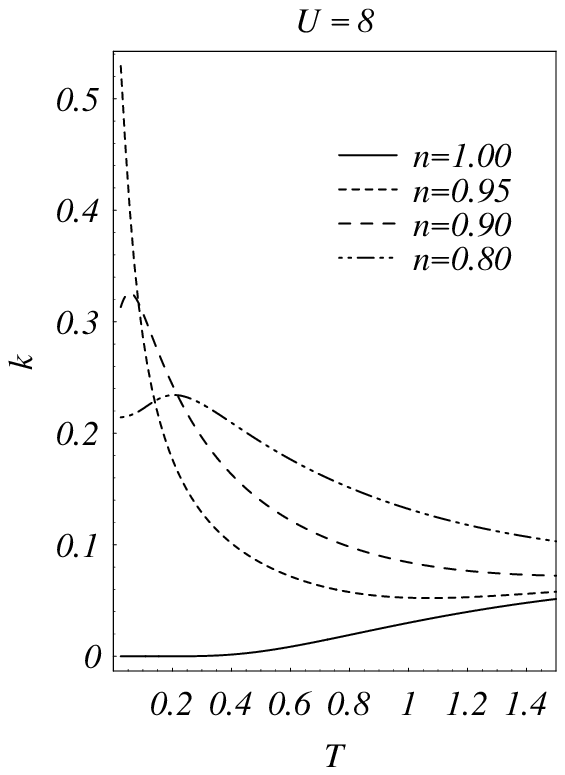}
    \includegraphics[width=0.31\textwidth]{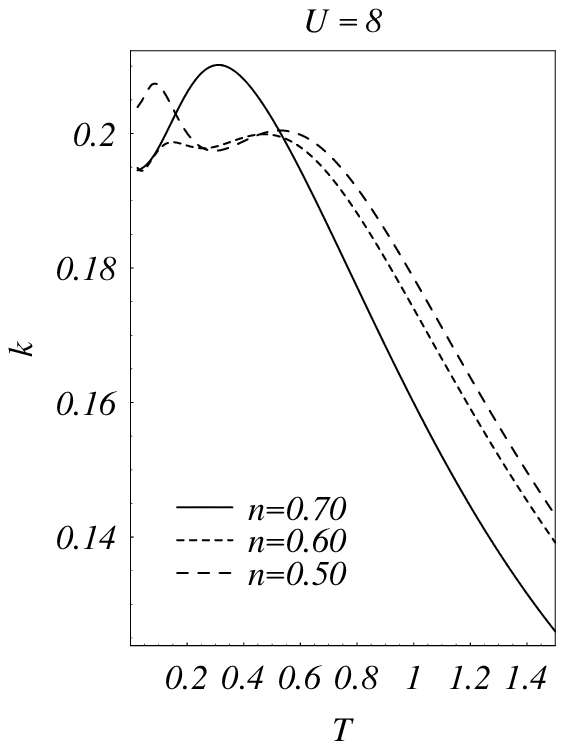}
    \includegraphics[width=0.31\textwidth]{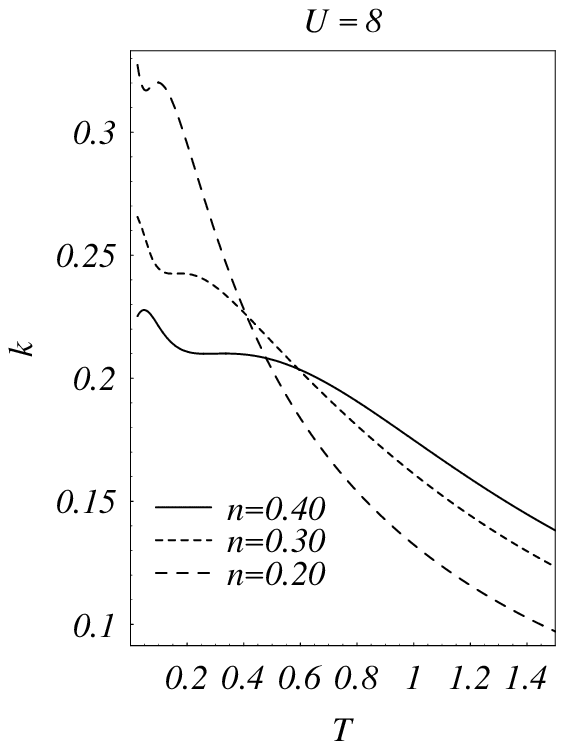}
  \end{center}
  \caption{Compressibility versus T for fixed $U=8$.}
  \label{fig:hubConstU-K}
\end{figure}
\begin{figure}[htb]
  \begin{center}
    \leavevmode  
    \includegraphics[width=0.32\textwidth]{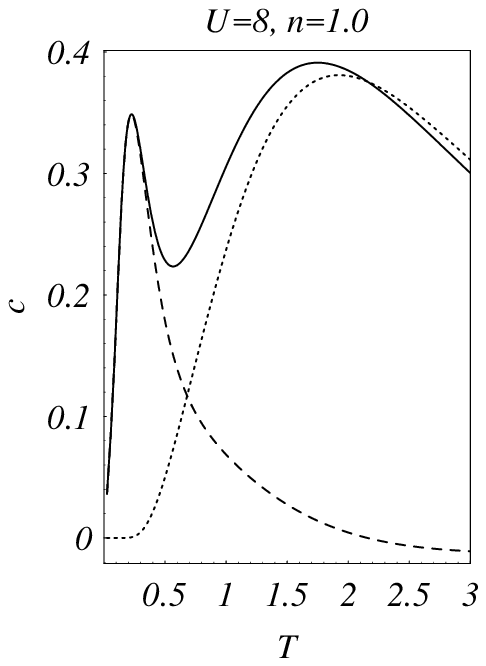}
    \includegraphics[width=0.32\textwidth]{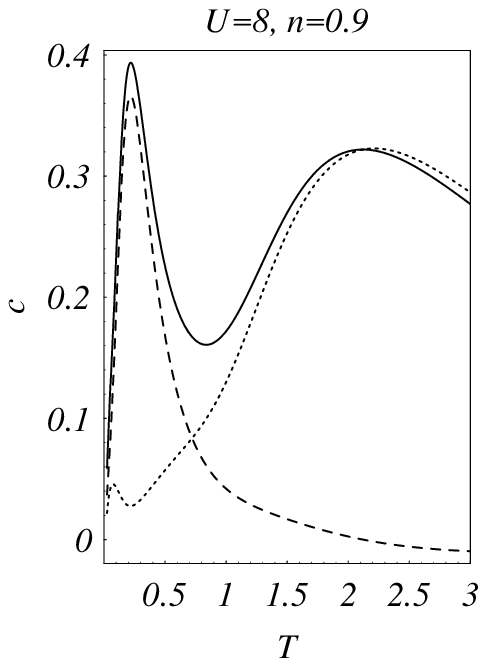}
    \includegraphics[width=0.32\textwidth]{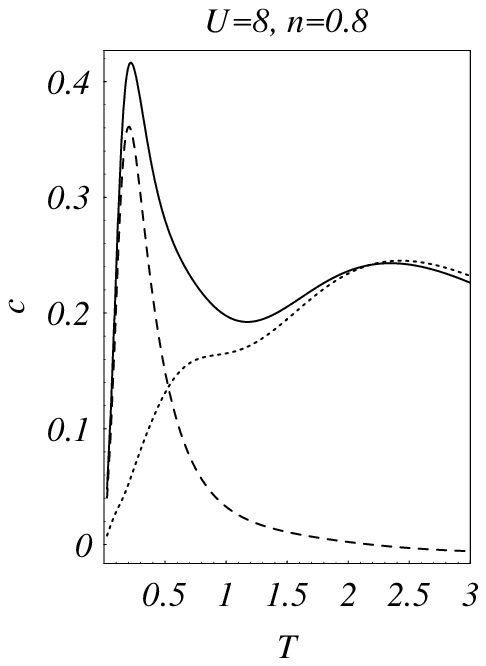}
  \end{center}
  \caption{Separation of specific heat (solid) in spin (dashed)
    and charge components (dotted).}
  \label{fig:hubSpinCharge}
\end{figure}

\end{document}